\documentclass[aps,reprint,showpacs,superscriptaddress,
footinbib,citeautoscript]{revtex4-1}

\usepackage{graphicx}

\usepackage[utf8]{inputenc}
\usepackage{csquotes}
\usepackage[american]{babel}
\usepackage[T1]{fontenc}
\usepackage{enumerate}
\usepackage{mdwlist}
\usepackage[activate=normal]{pdfcprot}
\usepackage{bbding}
\usepackage{color}
\usepackage{amssymb}
\usepackage{amsmath}
\usepackage{amsfonts}
\usepackage{mathrsfs}
\usepackage{lipsum}

\usepackage{bm}
\usepackage{dcolumn}
\usepackage{color}
\usepackage[colorlinks=true,citecolor=blue]{hyperref}
\hypersetup{colorlinks=true,citecolor=blue,linkcolor=red
,urlcolor=blue}

\begin{document}

\title{Thermoelectric transport properties of borophane }

\author{Moslem Zare}
\email{mzare@yu.ac.ir}
\affiliation{Department of Physics, Yasouj University, 75914-353, Yasouj, Iran}
\begin{abstract}
We theoretically study the influence of impurity scattering on the electric and thermal transport of borophane layer, a two-dimensional anisotropic Dirac semi-metal with two tilted and anisotropic Dirac cones.
In a systematic framework, we have calculated exactly the electrical conductivity and thermoelectric coefficients of borophane in the presence of the short-range, long-range charged impurity and the short-range electro-magnetic (SREM) scatterers, by using the exact solution of the Boltzmann transport equation within the linear-response theory.
Contrary to the large electron-hole asymmetry in borophane, its electron-hole conductivity is nearly symmetric.
Interestingly, for the short-range scatters, just like graphene, the short-range conductivities of borophane have the constant values, independent of the chemical potential, while the conductivities of the SREM scatterers are linearly dependent on the chemical potential.
Regardless of the impurity type, the electric conductivity of borophane is highly anisotropic, while the Seebeck coefficient and figure of merit (${\it ZT}$) are isotropic.
Along with the ambipolar nature of the borophane thermopower, a very high value of ${\it ZT}$ around unity is obtained at room temperature, due to the large asymmetry between electrons and holes in borophane.
More importantly, borophane attains its maximum value of ${\it ZT}$ at very low chemical potentials, in the vicinity of the charge neutrality point.
In comparison to phosphorene, a highly unique anisotropic 2D material, borophane with a higher anisotropy ratio ($\sigma_{xx}/\sigma_{yy}\sim10$), is an unprecedented anisotropic material. This high anisotropy ratio together with the large figure of merit, suggest that borophane is promising for the thermoelectric applications and transport switching in the Dirac transport channels.
\end{abstract}
\pacs{}
\maketitle

\section{Introduction}\label{sec:intro}

Over the past decades, boron, the left neighbor of carbon in the periodic table, has been expected to form various boron nanostructures, such as zero-dimensional all-boron fullerene-like cage cluster $B_{40}$~\cite{mannix2015synthesis,Gonzalez2007prl,Gonzalez2007nrl}, 1D boron nanowires and nanotubes~\cite{Otten2002,F.Liu-nanowire,Ciuparu2004,F.Liu2010}, double-ring tubular structures~\cite{Kiran2005,Oger2007,An2006} and 3D superhard boron phases~\cite{Eremets2001}, at the past decade. Among these mono-elemental structures graphene-like 2D structure of boron, known as borophene has attracted considerable attention both theoretically and experimentally, due to its exceptional properties and promising applications in nanoelectronics~\cite{H.Liu13,X.Wu12,Y.Liu13,Z.Zhang15}.

\par
Theoretical investigations indicate that due to the electrons in Fermi surface arising from the hybridized states of the $\sigma$ and $\pi$ bonds, 2D boron structure may be a pure single-element intrinsic superconducting material with the highest $T_c$ (higher than the liquid hydrogen temperature), on conditions without high pressure and external strain which can be modified by strain and doping~\cite{Penev-NL16,M.Gao,RC.Xiao}.

Several types of borophene have been synthesized on Ag(111)~\cite{mannix2015synthesis,ZZhang,BFeng,BFeng2}. A similar striped phase, named as $\beta_{12}$ borophene, has an essentially flat structure that weakly interacts with the Ag(111) substrate. In particular, the existence of the Dirac cones with ultrahigh Fermi velocity, in $\beta_{12}$ borophene clearly proved using the angle-resolved photoemission spectroscopy (ARPES) experiment as well as by first-principles calculations~\cite{Boro}.

\par
This novel two-dimensional material was first proposed by Boustani, who predicted that the quasi-planar 2D boron sheet could be constructed from a basic unit of puckered $B_7$ cluster by using the systematic ab initio density functional method $B_7$ cluster~\cite{Boustani1,Boustani2}.
The first experimental evidence of the atomic-thin boron sheets was performed by Piazza { \it et al.} in the planar $B_{36}$ cluster with a central hexagonal hole~\cite{Piazza14}.

As proposed by Tang and Ismail-Beigi, the mixed hexagonal-triangular 2D boron sheet ($\alpha$-borophene) was thought to be more stable than one composed only of buckled triangular motifs~\cite{Tang2007,Tang2009,Tang2010}.
Using the first-principle particle-swarm optimization global algorithm, Zeng’s group showed that another two flat monolayers as $\alpha_1$-borophene and $\beta_1$-borophene, are energetically most stable 2D boron structures among the state-of-the-art 2D structures of boron~\cite{X.Wu2012}.

\par
Recently, a new type of 2D boron Polymorph with an orthorhombic 8-$ Pmmn $ symmetry group, has been predicted to be more stable than the $\alpha$-borophene, that exhibits anisotropic tilted Dirac cones~\cite{zhou2014semimetallic}. A first-principles study reveals that borophene is the first known materials with high-frequency plasmons in the visible spectrum~\cite{Sadrzadeh-NL12}.
Furthermore, in this borophene polymorph the anisotropic plasmon modes remain undamped for higher energies along the mirror symmetry direction, in which the anisotropic Friedel oscillation behaves like $r^{-3}$ in the large-$r$ limit~\cite{Sadhukhan2017}.

\par
However, theoretical calculations show that due to the imaginary frequencies in its phononic dispersion, the Free-standing 2D borophene ($B_8$) is unstable against long-wavelength periodic vibrations~\cite{Xu16,Jena2017}, needing a substrate to be stabilized. A feasible and effective method to dynamically stable borophene, is the chemical functionalization using surface hydrogenation. First-principles calculations of Xu {\it et al.} show that the fully hydrogenation of borophene, called borophane ($B_2H_2$), is a viable method to stabilize borophene in the vacuum without a substrate~\cite{Xu16}. It has also been predicted in this work that borophane has a remarkable Fermi velocity which is nearly four times higher than that of graphene.

An {\it et al.} using first-principles density functional theory plus the non-equilibrium Green’s function approach, demonstrated that borophane displays a huge electrical and mechanical anisotropy. Along the valley-parallel direction (armchair), the 2D borophane exhibits a metallic treatment with a linear current-voltage curve, in contrast to the perpendicular buckled direction (zigzag), that shows off a semiconductor behavior ~\cite{Anpccp2018}.

Similar to hydrogenated graphene, borophene is capable of forming various two-dimensional allotropes due to its valency. Only one case has been reported in literature as far as our knowledge, the chair-like borophane, named C-boropane.
It is intriguing to know, whether there are other metastable allotropes with novel electronic characteristics?

Very recently, in a comprehensive study~\cite{WangSci16}, the likely formation of borophane with a focused on the structural stability of borophane allotropes, has been provided, based on the first principles calculations. The crystal structure of seven allotropes of borophane, namely Plane-Square-type borophane (PS-borophane), Plane-Triangle-type borophane (PT-borophane), Chair-like borophane (C-borophane), Boat-like borophane (B-borophane), Twist-Chair-Boat-type borophane (TCB-borophane), Triangle-type borophane (T-borophane) and Washboard-like borophane (W-borophane), were identified in this paper~\cite{WangSci16}. The results show that the charge transfer from B atoms to H atoms is crucial for stability of an borophane allotrope. W-borophane, the most stable allotrope of the hydrogenated borophene, has energy about 113.41 meV per atom, lower than that of C-borophane ~\cite{WangRSC17}.

\par
Thermoelectric materials, based on a fundamental interplay between their electronic and thermal performance, have attracted much interest for energy efficient device applications~\cite{Dresselhaus93, Harman, Venkatasubramanian, Arita, Hamada, Zide, Wei, Zuev, Kato, Buscema, Konabe1}. The efficiency of the refrigeration or power generation devices is characterized by a dimensionless figure of merit ${\it ZT}=\frac{\sigma\mathcal{S}^{2}}{\mathcal{K}}T$, in which $\sigma,\mathcal{K}$ and $\mathcal{S}$ are electrical conductivity, thermal conductivity and Seebeck coefficient (thermopower), respectively and $T$ is absolute temperature. The small thermal conductivity and relatively high thermopower and electrical conductivity are required for high-efficiency thermoelectric materials. Even if the Seebeck coefficient becomes large, a heat current inevitably accompanies a temperature gradient and thus makes a tradeoff.

Therefore, a great deal of efforts has been carried out to improve the thermoelectric performance through balancing these interdependent thermoelectric parameters. There are completely different routes to improve the thermoelectric figure of merit, which are tailoring to improve the power factor ($\sigma\mathcal{S}^{2}$) and lowering the thermal conductivity.

A hallmark of the thermoelectric materials study came actually from the pioneering work~\cite{Dresselhausprb47-93-1}, proposing that nanostructuring materials may enhance thermoelectric efficiency, due to the sharp-peaked electronic density of states (DOS) in low-dimensional materials~\cite{Dresselhausprb47-93-1, Bilc}. The Seebeck coefficient, which depends on logarithmic derivative of DOS, is significantly enhanced in 2D materials and hence, the thermoelectric efficiency increases ~\cite{Mahan}.
This seminal work of the thermoelectric materials could be an important starting point for today's achievements. Due to the decreased thermal conductivity caused by phonon boundary scattering and consequently, improved figure of merit on quantum confinements in low dimensional systems, ${\it ZT}$ values become dramatically larger than the corresponding bulk materials.

Using energy filtering proposal, the Seebeck coefficient increases by introducing a strongly energy-dependent scattering mechanism
~\cite{Mahan, Neophytou, Fomin, Zianni}.
Other useful methods are also used for enhancing thermoelectric performance. The use of band structure engineering~\cite{Heremanssci08,Liu-Zhao-APL08} in conjunction with nanostructuring, to lower the thermal conductivity, could further enhance figure of merit of thermoelectric materials.
It is known that quantum confinement of carriers in bulk samples, containing nanostructured constituents~\cite{Dresselhaus07}, quantum-well super-lattices ~\cite{Dresselhausprb47-93-1}, as well as quantum-wires~\cite{Dresselhausprb47-93-2}, will enhance thermoelectric efficiency in order to promote the more widespread use of thermoelectric materials.

According to a comprehensive review on the thermoelectric materials~\cite{Xiao-npj18}, among the well-known thermoelectric materials, PbTe is distinguished as a very promising compound for power generation at intermediate-temperature range (500-900 K), successfully served in several NASA space missions~\cite{LaLonde}.
At near room temperatures (300–500 K), appropriate thermoelectric candidates are (Bi, Sb)$_2$(Se,Te)$_3$-based alloys and MgAgSb alloys, while at high temperature (>900 K), half-Heusler (HH) alloys, (Pr, La)$_3$ Te$_4$, SiGe, and Yb$_{14}$ Mn Sb$_{11}$, are very promising candidates~\cite{Xiao-npj18}.
Another important thermoelectric materials are doped narrow-gap semiconductors~\cite{Arita, Hamada, Zide}, PbTe(1.5$~$nm)/Pb$_{0.927}$Eu$_{0.073}$Te(45~nm), multiple quantum well~\cite{Dresselhausprb47-93-1, Harman} and Bi$_{2}$Te$_{3}$~\cite{Venkatasubramanian}.

Recent advances in fabrication technologies have made exploring two-dimensional materials possible for thermoelectric applications~\cite{Wei, Zuev, Kato, Buscema, Konabe1}.

As already pointed out, semi-metals with large electron-hole asymmetry can be considered as an important strategy for strong enhancement of the thermoelectric coefficients~\cite{Markov-scr18}. Moreover, motivated by the great interest in search of thermoelectric Dirac materials, the knowledge of thermoelectric efficiency of borophane is crucial for its application in potential thermal management devices. However, to our best knowledge, answers to these question is still lacking.

\par
In this paper, with a view to investigate the effect of various impurities on the thermoelectric performance of borophane, a Dirac semi-metals with a large electron-hole asymmetry, we perform a systematic study on the electrical and thermal transport in borophane, by means of the Boltzmann transport equation.
We first propose an accurate low-energy model Hamiltonian for borophane, then, the electronic contribution to the thermoelectric transport of the monolayer borophane is investigated. We consider a monolayer borophane in diffusive transport regime in the $x$–$y$ plane, driven by a lattice temperature gradient ${\vec{\nabla }}T$ and an electric field ${\vec{\mathcal{E}}}$. The generalized Boltzmann transport equation is applied to obtain the conductivity, Seebeck coefficient and the figure of merit, by considering the various types of resistive scattering potentials, (1) the short-range (SR) potential, (2) the long-range (LR) Coulomb potential, with a Thomas- Fermi screening as the source of scattering and (3) the impurities containing short-range electric and ferromagnetically ordered magnetic potentials, called electro-magnetic (SREM) scatterers. At low temperature elastic scattering becomes the dominant mechanism as inelastic scattering is strongly suppressed. In this work our focus is on elastic scattering and therefore, with good approximation, intervalley scattering (interband processes) is neglected ~\cite{E.H.Hwang08}.

\par
Our calculations show that although the electrical conductivity of borophane is highly anisotropic, but the Seebeck coefficient and the corresponding figure of merit, irrespective of the underlying scattering mechanisms, are isotropic. Interestingly, for the short-range scatters, just like graphene, the short-range conductivities of borophane have constant values (independent of the chemical potential), that are $\sigma_{xx}=250 \ e^2/h$ and $\sigma_{yy}=33 \ e^2/h$, at $T=20\, {\rm K}$, along the armchair and zigzag direction, respectively. As a measure of thermoelectric efficiency, the figure of merit, reaches to about $2.75$ and $1$, at $T\sim20 \, {\rm K}$ and room temperature ($T=300\, {\rm K}$), respectively. We also investigate the effect of the magnetic scatters on thermoelectric transport coefficients. The conductivities of the SREM scatterers have a linear dependence on the doping levels. These results propose that borophane could be a promising material for the thermoelectric applications.

\par
The paper is organized as follows. In Sec.~\ref{sec:model}, at first the model Hamiltonian and lattice structure of borophane is introduced and then the method which is used to calculate the conductivity and thermoelectric coefficients using the generalized Boltzmann method, is explained. In Sec.~\ref{sec:results}, we present and describe our numerical results from the exact calculations for the conductivity and thermoelectric coefficients for borophane. Finally, we conclude and summarize our main results in Sec.~\ref{sec:concl}.

\section{Model Hamiltonian of borophene}\label{sec:model}
\begin{figure}
\centering
\includegraphics[width=3.3in]{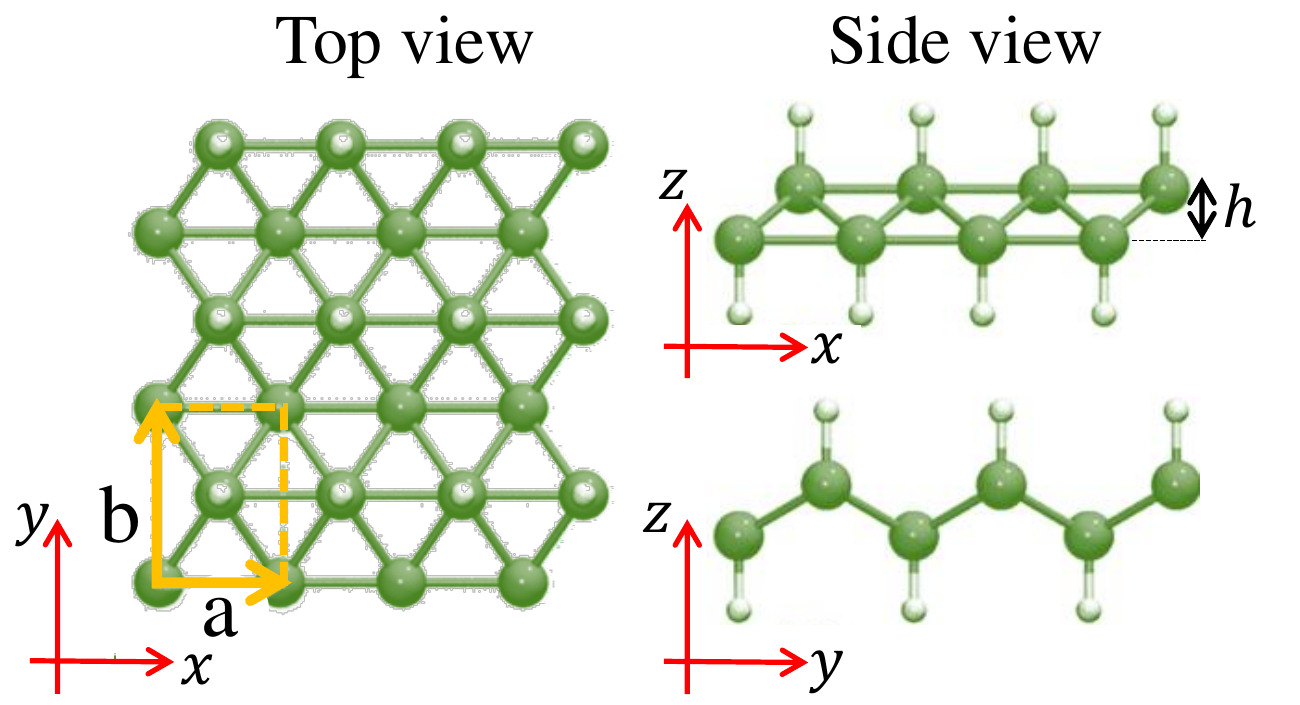}
\caption{(Color online) Schematic drawing of the optimized ground-state structure of borophane lattice, with top (left panel) and side view (right panel). The unit cell is marked with a yellow rectangle, contains two boron (B) atoms and two hydrogen (H) atoms and the basic vectors of the primitive unit cell are indicated by the yellow arrows. The green and white balls represent B and H atoms, respectively. The buckling height is denoted by h.}
\label{lattice}
\end{figure}

In the following sections, we consider a monolayer of borophane at low temperature. The Bravais lattice constants of the conventional orthorhombic unit cell (contains  four atoms, as seen in Fig.~\ref{lattice} ) of the buckled crystal structures of borophene and borophane, are $a = 1.62$ \AA~, $b = 2.85$\AA~ and $a = 1.92$ \AA~, $b= 2.81$ \AA~, respectively~\cite{Piazza14,Xu16,Tang2007}. Notice that the buckling height of $h= 0.96$ \AA~ in borophene reduces to $h= 0.81$ \AA, upon hydrogen adsorption in borophane~\cite{Padilhapccp16}.

In combination with first-principles calculations and starting from the simplified linear combination of atomic orbitals approach , M. Nakhaee {\it et al.} have constructed a tight-binding model in the two-centre approximation for borophene and borophane~\cite{Peeters-Hamilton}. The Slater and Koster approach is applied to calculate the TB coefficients of these two systems.

Our starting point is the low-energy continuum Hamiltonian for borophane ~\cite{Peeters-Hamilton}, that describes an anisotropic and tilted Dirac crossing along the $ \Gamma $-X direction in the rectangular Brillouin zone. In the vicinity of two nonequivalent Dirac points ${\bf K}_D=(\pm 0.64,0)$\AA$^{-1}$ ~\cite{zabolotskiy2016strain,Peeters-Hamilton,Sadhukhan2017,Ezawa2017,Islam2017}

\begin{equation}
H = \hbar v_{x} k_{x}\sigma_{x} + \hbar v_{y}k_{y} \sigma_{y} + \hbar v_{t}k_{x} \sigma_{0}.
\label{boeqn}
\end{equation}

Here, $\sigma_{x},\sigma_{y} $ are the Pauli matrices for the pseudospin representing the lattice degree of freedom while $ \sigma_{0} $ is the $ 2 \times 2 $ identity matrix. Typical values of the direction-dependent velocities, in units of $\left(\times 10^{5}\,m/s \right)$, are $ v_{x} = 19.58 $, $ v_{y} = 6.32 $, and $ v_{t} = -5.06 $. The corresponding energy dispersion of the Hamiltonian Eq. (\ref{boeqn}), is analytically given by
\begin{equation}
E^{\tau}\left(k\right) = \hbar v_{t}k_{x}+\tau \hbar\sqrt{v_{x}^{2}k_{x}^{2} + v_{y}^{2}k_{y}^{2}},
\label{diseq}
\end{equation}

in which $\tau=$1(-1) denotes the conduction (valence) band in borophane dispersion. Isoenergy contour map of the electronic band structure of borophane in the $k-$space for $E(\vec{k}_{\rm F})=-0.5$ to $0.5$ eV, is shown in Fig.\ref{energyDispersion}(a).

\par
Since several works on electronic properties of the 8-$ Pmmn $ borophene and borophane are available, a proper comparison with those results seems to be in order. Recent investigations, based on an {\it ab initio} evolutionary structure calculations~\cite{X.-F.Zhou2016}, show that the Fermi velocity value of the Dirac fermions in borophene is $v_F^x=0.56\times 10^{6}\,m/s $ in the $k_x$ direction, while for the $k_y$ direction it has two values of $v_F^y=0.46 \times 10^{6}\,m/s $ and $1.16 \times 10^{6}\,m/s $. For borophane, in the $k_x$ direction, the predicted Fermi velocities are $v_F^x=1.74$ and $0.97$, both in units of ($10^{6}\,m/s $). These values of the Fermi velocity are reported also in Ref.~\cite{zhou2014semimetallic}. It is worthwhile to mention that the predictive values of velocities $v_F^x, v_F^y$, were evaluated via the slope of the bands in the $k_x$ and $k_y$ directions, respectively by the formula $v_F=E(k)/\hbar k$.

In a comprehensive study, Feng \textit{et al.} ~\cite{Boro} based on tight-binding analysis revealed that the Fermi velocities in borophene are approximately $v_F^x=6.1$ eV and $v_F^y=7.0$ eV \AA, which are close to the Fermi velocity of graphene ($v_F=6.6$ eV \AA). These values also confirmed by angle-resolved photoemission spectroscopy and first-principles calculations~\cite{Boro}.

As previously mentioned, the first principle calculations show that the same pair of Dirac cones, is revealed for $\beta_{12}/Ag(111)$~\cite{Boro}. The calculated Fermi velocity for $\beta_{12}/Ag(111)$ is approximately $v_F=3.5$ eV \AA, which is in the same order of magnitude as the experimental value. The difference between the theoretical and experimental values might originate from the many-body interactions~\cite{Boro}.

A cat’s-cradle-like topological semimetal phase which looks like multiple hourglasslike band structures staggered together, was discovered by Fan \textit{et al.}~\cite{X.Fan2018}. An original effective Hamiltonian based on $k.p$ approximation was extracted for monolayer borophene and borophane, the first material class to realize such a semimetal phase, including the SOC effects.
In this work, with a comprehensive investigation of topological properties  ~\cite{X.Fan2018}, by fitting the $k.p$ model with the DFT results, the corresponding Fermi velocities for Dirac fermions are evaluated to be $v_F^x=8.0$, $v_F^y=5.2$, $v_t=3.4$, for monolayer borophene and $v_F^x=13.9$, $v_F^y=7.7$, $v_t=3.5$, for monolayer borophane, respectively, all in units of ($10^{5}\,m/s $).

Zabolotskiy \textit{et al.} ~\cite{zabolotskiy2016strain} have developed a tight-binding model Hamiltonian for a 2D Dirac semimetal 8-$ Pmmn $ borophene, evaluated in Ref.~\cite{zhou2014semimetallic}. In this work, the Hamiltonian parameters were found from a fit to the DFT data and the resulting electronic structure reproduces the main features of the DFT band structure, including the Dirac points at the Fermi level.
Within their model Hamiltonian, the velocities $v_x,v_y,v_{t}$ in vicinity to the K point are given by the following expressions

\begin{eqnarray}
&v_\mathrm{t} = -\frac{\partial^2P/\partial k_y\partial E}{\partial^2P/\partial E^2},\nonumber\\
v_x^2 &=-\frac{\partial^2P/\partial k_x^2}{\partial^2P/\partial E^2},\quad
v_y^2 = v_\mathrm{t}^2-\frac{\partial^2P/\partial k_y^2}{\partial^2P/\partial E^2};\\
&\frac{\partial E}{\partial k_x}=\pm v_x,\quad \frac{\partial E}{ \partial k_y}=\pm v_y+v_\mathrm{t}. \label{slopes}
\label{}
\end{eqnarray}

 where $P=P(E,\mathbf{k})$ is the characteristic polynomial for energy $E$ and momentum $\mathbf{k}$, and $E=E(\mathbf{k})$ is the eigenvalue.
The three velocities are given by $\{v_x,v_y,v_\mathrm{t}\}$=$\{0.86,0.69,0.32\}$ $\times10^6  m/s$  ~\cite{zabolotskiy2016strain}.

Cheng \textit{et al.} ~\cite{T.Cheng17}, developed an effective Hamiltonian around the Dirac point, to a linear term of the wave vector, and obtained an analytical formula for prediction of the intrinsic carrier mobility in 2D materials with tilted Dirac cones and combine it with first-principles calculations to determine the properties of semimetallic borophene and borophane.
Their predicted velocities in  8-$ Pmmn $ borophene are $v_F^x=7.85$, and $v_F^y= 5.34$, and for monolayer borophane, they are  $v_F^x=13.48$, $v_F^y= 7.7$, along the x and y directions, respectively, and the tilting velocities are  $v_t=-3.45$, and $v_t=-3.86$ in 8-$ Pmmn $ borophene and borophane, respectively. Note that the all velocities are in units of ($10^{5}\,m/s $).

\begin{figure}
\centering
\includegraphics[width=3.3in]{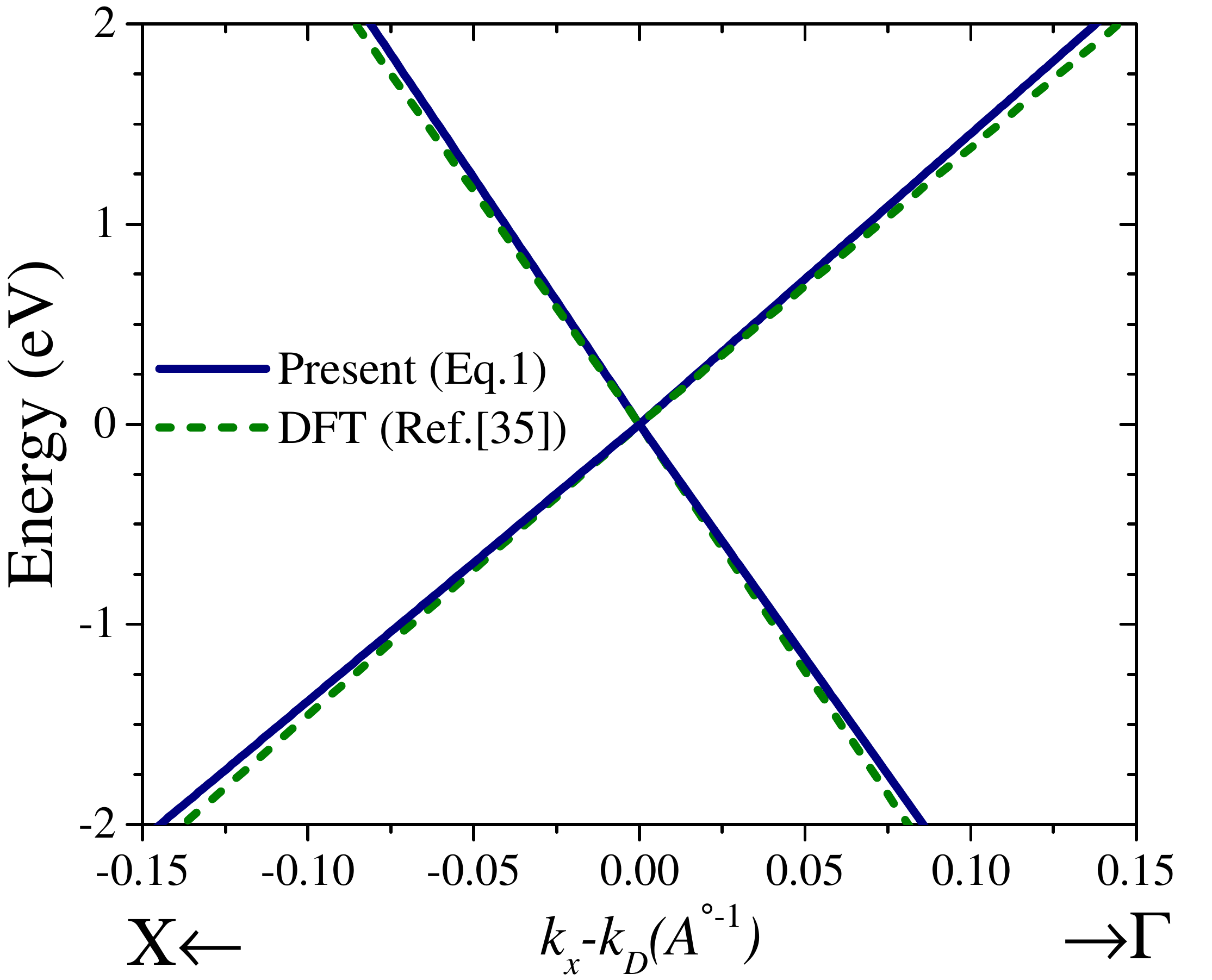}
\caption{(Color online)  The band structure of borophane in the first Brillouin zone along the $\Gamma-X$ direction. We have compared our used model Hamiltonian (Eq.\ref{boeqn}) with the calculations, based on the density functional theory and non-equilibrium Green’s function approaches~\cite{Anpccp2018}.}\label{fig-dis}
\label{dispersion}
\end{figure}

\begin{figure}
\centering
\includegraphics[width=3.5in]{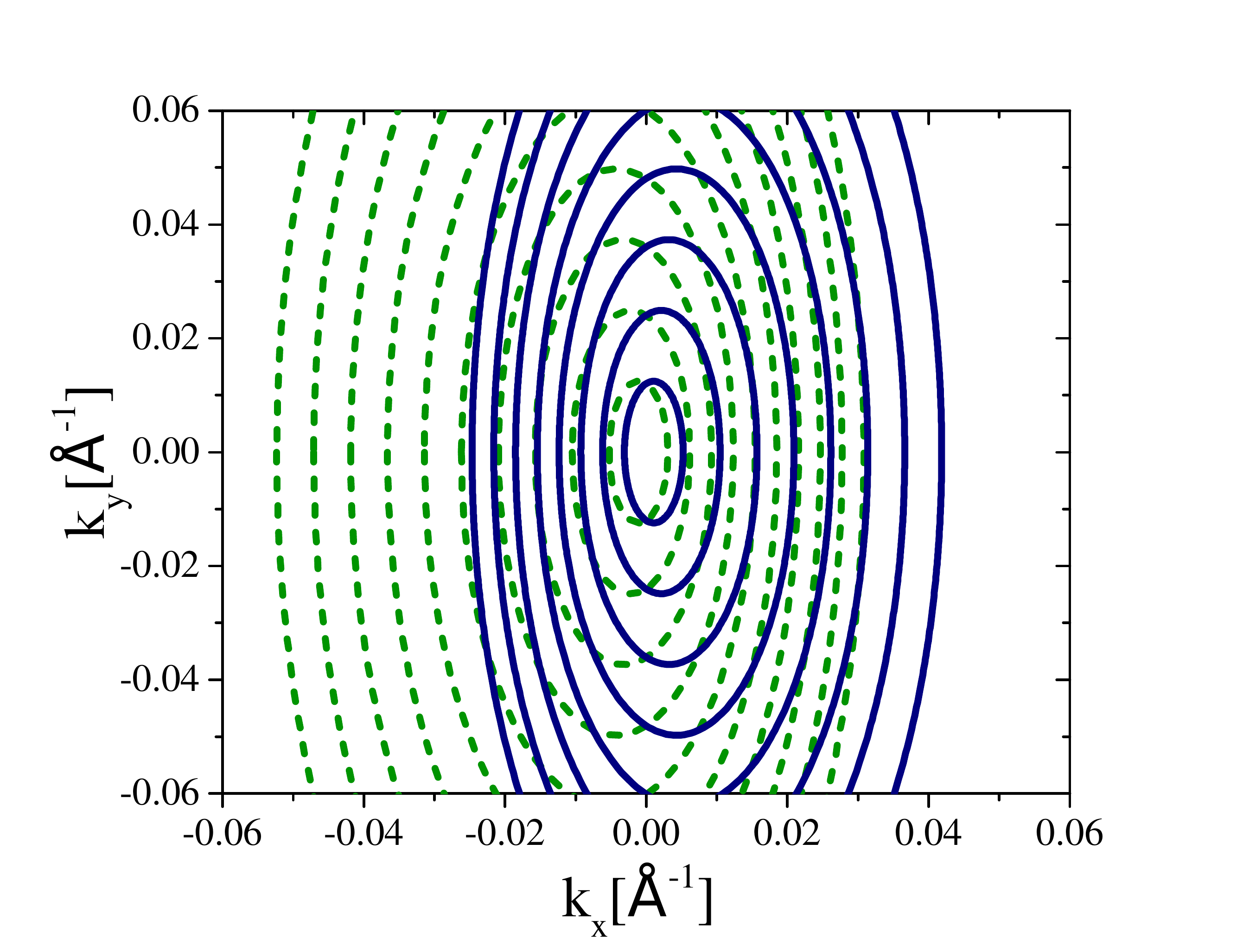}
\includegraphics[width=3.5in]{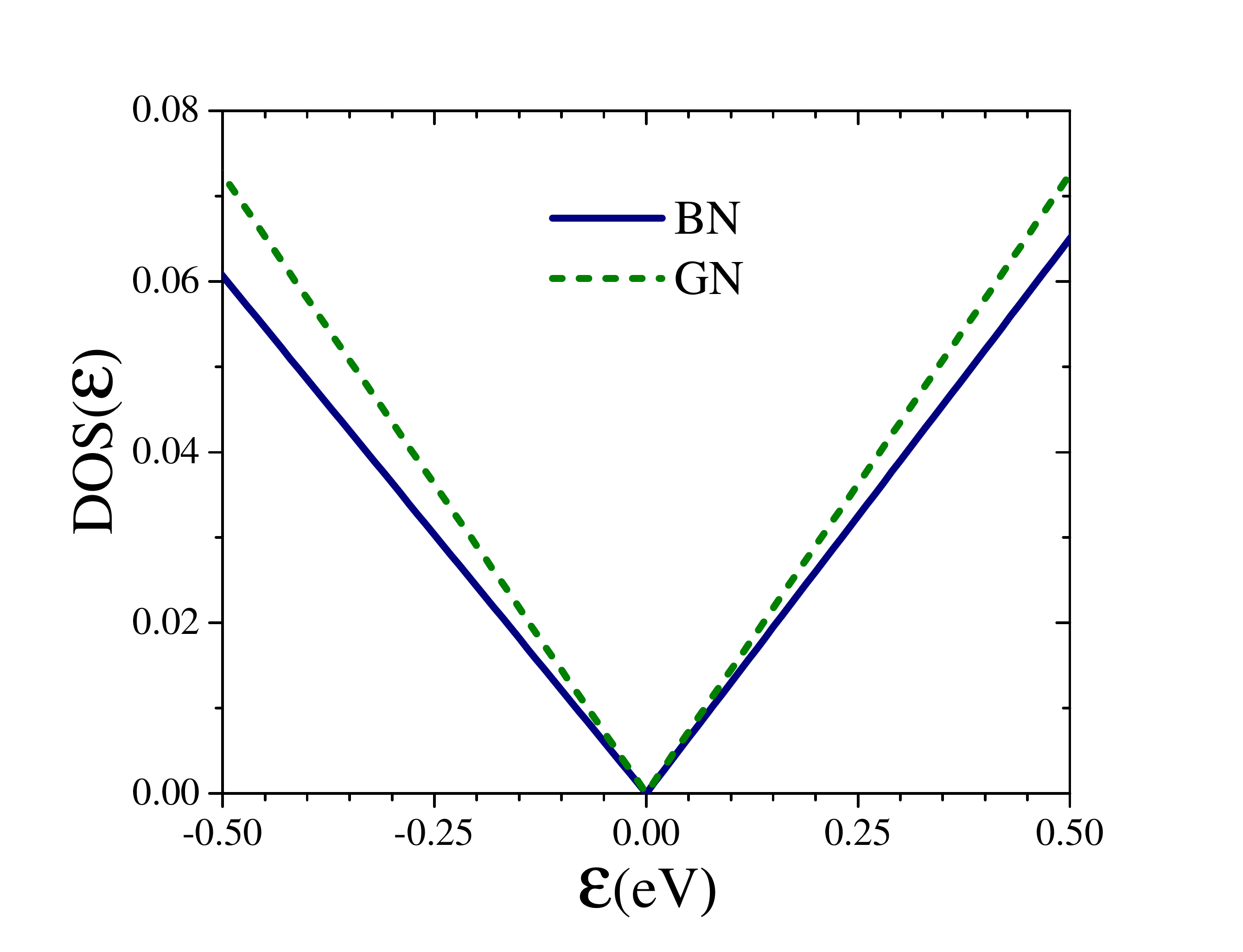}
\caption{(Color online) (a) Isoenergy contour map of the electronic band structure of borophane in the $k-$space for
$E(\vec{k}_{\rm F})=-0.5$ to $0.5$ eV. The contour lines are drawn at $0.05$ eV intervals in both the electron(solid-navy) and hole (dashed-green) doped cases. (b) Density of states (DOS) calculated for monolayer borophane (solid-navy) and graphene (dashed-green).}
\label{energyDispersion}
\end{figure}

The contour lines are drawn at $0.05$ eV intervals in both electron (solid-navy) and hole-doped cases (dashed-green).We demonstrate the band dispersion of borophane (Eq.\ref{diseq}) in Fig.~\ref{dispersion}, where we have compared our used model Hamiltonian (Eq.\ref{boeqn}) with that calculations, based on the density functional theory and non-equilibrium Green’s function approaches~\cite{Anpccp2018}, the so-called NEGF-DFT method.

The corresponding eigenfunction for the Dirac electrons is given by the following equation

\begin{equation}\label{wave}
\psi^{\tau}({\vec{k}})=\frac{1}{\sqrt{2A}}\left[\begin{array}[c]{c}
                                   \tau e^{-i\beta_{\vec{k}}}\\
                                   1
                                  \end{array}\right]e^{i\vec{k}\cdot\vec{r}},
\end{equation}
where $A$ is the system area and $\beta_{k}=\tan^{-1}\left[ v_y k_y/v_x k_x \right]$.

Furthermore, By invoking the band energy dispersion given by Eq. (\ref{diseq}), the $x$ and $y$ components of the velocity can be calculated as
\begin{eqnarray}
v_{x}^{\tau}(k)&=&v_{t}+\tau \frac{v_x^{2}k_x}{\sqrt{v_x^{2}k_x^2+v_y^{2}k_2^2}}\\
v_{y}^{\tau}(k)&=&\tau \frac{v_y^{2}k_y}{\sqrt{v_x^{2}k_x^2+v_y^{2}k_2^2}}
\end{eqnarray}

The density of states (DOS) can be obtained by solving the following equation

\begin{eqnarray}\label{eq-54}
 D(\varepsilon) &=& \displaystyle \frac{1}{(2\pi)^2} \int_0^\infty k'\, dk'
  \delta(\varepsilon-\varepsilon_{\vec{k}'})
\end{eqnarray}

Performing this integral over energy, one finds the following expansion for the density of states:

\begin{eqnarray}\label{eq-54}
  D(\varepsilon) =  \frac{k^{\tau}(\varepsilon,\phi) }{v_t\cos(\phi) + \sqrt{v_x^2 \cos^2(\phi)+ v_y^2 \sin^2(\phi)}}
\end{eqnarray}

where $\phi=\tan^{-1}(k_y/k_x)$ and the wave vector $ k^{\tau}(\varepsilon,\phi)$ is given by

\begin{widetext}
\begin{eqnarray}\label{eq-54}
 k^{\tau}(\varepsilon,\phi) =\varepsilon  \frac{-2 v_t\cos(\phi) +\tau \sqrt{2}\sqrt{ v_x^2 +  v_y^2 +  (v_x^2-  v_y^2) \cos(2\phi)}}{v_x^2 + v_y^2 - v_t^2 + (v_x^2 - v_y^2 - v_t^2) \cos(2\phi)}
\end{eqnarray}
\end{widetext}

Interestingly, similar to the low-energy DOS of graphene, DOS of borophane has the well-known linear form, as shown in Fig.\ref{energyDispersion}(b).
In this figure, also the DOS of graphene (dashed-green) is plotted by setting $v_x=v_y=1\times10^{6}\,m/s $ and $v_t=0$.

\subsection{Anisotropic transport framework}

In this section, we present our main formalism for the numerical evaluation of the thermoelectric coefficients of borophane, in the diffusive regime, using the generalized semiclassical Boltzmann equation.
In particular, we take into account three important cases of short-range impurities with Dirac delta potential (\textit{e.g.}, defects or neutral adatoms), the long-range Coulomb impurities and finally the electro-magnetic scatterers, containing short-range electric and ferromagnetically ordered magnetic potentials. Electrical conductance, thermal conductance, Seebeck coefficient, and thermoelectric figure of merit ({\it ZT}) of borophane in the presence of both the electric field and the temperature gradient will be studied.

In order to calculate the transport coefficients, we use the following expression for the charge current ${\bf j}$ and energy flux density ${\bf j}^{q}$
\begin{eqnarray}
\left[\begin{array}{c}
{\bf j}  \\
{\bf j}^{q}
\end{array}
\right]=\int\frac{d^2k}{(2\pi)^2}
\left[\begin{array}{c}
-e  \\ \varepsilon({\vec{k}})-\mu
\end{array}
\right]{ v(\vec{k})} f({\vec{k}})
\label{currents}
\end{eqnarray}
where ${ v(\vec{k})} =v(\phi)(\cos\xi, \sin\xi)$ is the semiclassical velocity of the carriers, which is related to the energy dispersion
$\varepsilon_{\vec{k}}$ through ${\bf v}=(1/\hbar)\nabla_{\vec{k}}\varepsilon_{{\vec{k}}}$ . $f({\vec{k}})$, is the nonequilibrium quantum distribution function, describes the evolution of the charge distribution in the presence of thermoelectric forces, must be computed from the Boltzmann equation. $\xi$ is the angle of the velocity vector with respect to $x$ axis.
For this purpose, we take the Boltzmann equation up to a linear order in the presence of thermoelectric forces

\begin{eqnarray}
&&\hspace*{-10pt}\left(-e{\vec{\mathcal{E}}}+\frac{\varepsilon-\mu}{T}{\vec \nabla} T\right)\cdot{\vec{v}}({\vec{k}})\left[-\partial_{\varepsilon}f^{0}(\varepsilon_{{\vec{k}}})\right]=\left(\frac{df}{dt}\right)_{{\rm coll.}}
\label{2dani}
\end{eqnarray}
where $w({\vec{k}},{\vec{k}'})$ is the scattering rate from state ${\vec{k}}$ to state ${\vec{k}'}$ which needs to be specified according to the microscopic origin of the scattering mechanisms and $f^{0}$ is the equilibrium distribution function.

The collision integral is given by
\begin{eqnarray}
\left(\frac{df}{dt}\right)_{{\rm coll.}}=\int\frac{d^2k'}{(2\pi)^2}w({\vec{k}},{\vec{k}'})\left[f({\vec{k}},{\vec{\mathcal{E}}}, T)-f({\vec{k}'}, {\vec{\mathcal{E}}}, T)\right]\nonumber\\
\label{eq4}
\end{eqnarray}

Using the relaxation time approximation, the nonequilibrium distribution function cannot be exactly calculated and provides an inadequate explanation for the full aspects of the anisotropic features of the transport properties.

In the next section, we have implemented an approach for finding the exact solution to the linear-response Boltzmann equation for two dimensional anisotropic systems~\cite{Vyborny, bzare2016, Faridi,moslem-bp2}. Within the lowest order of the Born approximation, the scattering rates $w({\vec{k}},{\vec{k}'})$, using the Fermi golden rule are given by
\begin{eqnarray}
w({\vec{k}}, {\vec{k}'})=\frac{2\pi}{\hbar}n_{{\rm imp}}\big\vert\langle {\vec{k}'}|\hat{V}|{\vec{k}'}\rangle\big\vert^2 \delta(\varepsilon_{\vec{k}}-\varepsilon_{\vec{k}'})
\end{eqnarray}

where $n_{{\rm imp}}$ is the background random-charged impurity density and $\hat{V}_{{\vec{k}}-{\vec{k}'}}$ is the Fourier transformation of the interaction potential between an electron and a single impurity.

The short-ranged impurities are approximated with a zero-range hard-core potential $\hat{V}_{{\vec{k}}-{\vec{k}'}}=V_{0}$, while the long-ranged electron-electron Coulomb potential, owing to the charged impurities, is screened by other electrons of the system, according to the Thomas-Fermi approximation.

By invoking the expression for $f(\theta, \phi)$ into Eq. (\ref{currents}) for the charge and heat currents, the response matrix, which relates the resulting generalized currents to the driving forces, can be expressed in terms of some kinetic coefficients $\mathcal{L}^{\alpha}$ as the following \cite{moslem-bp2},
\begin{eqnarray}
\begin{pmatrix}
{\bf j} \\  {\bf j}^{q}
\end{pmatrix}=
\begin{pmatrix}
\mathcal{L}^{0} & -\mathcal{L}^{1}/eT\\
\mathcal{L}^{1}/e & -\mathcal{L}^{2}/e^{2}T
\end{pmatrix}
\begin{pmatrix}
{\vec{\mathcal{E}}} \\ -{\vec{\mathcal{\nabla}}} T
\end{pmatrix}
\label{eqbolt}
\end{eqnarray}

in which the two diagonal terms in coefficients matrix are the electrical $\sigma$ and thermal $\mathcal{K}$ conductivities, and the two off-diagonal elements  are the mixed coefficients, relating electrical and thermal phenomena through the Onsager's reciprocity relations. The thermoelectric power (or Seebeck coefficient) $\mathcal{S}=-\frac{1}{eT}(\mathcal{L}^{0})^{-1}\cdot\mathcal{L}^{1}$, describes the voltage generation due to the temperature gradient while Peltier coefficient $\Pi=T\mathcal{S}$ accounts for the heat current induction due to the charge current, respectively.
The ability of a given material to efficiently produce thermoelectric power is related to its dimensionless figure of merit. By parameterizing ${\vec{\mathcal{E}}}$ and $\vec{k}$, as ${\vec{\mathcal{E}}}=\mathcal{E}(\cos\theta, \sin\theta)$, ${\vec k}=k(\cos\phi, \sin\phi)$, respectively, all of the coefficients obey the relation
\begin{eqnarray}\label{eq:L}
\mathcal{L}^{\alpha}(\theta, \theta')=\int d\varepsilon\left[\frac{-\partial f_{0}}{\partial\varepsilon}\right](\varepsilon-\mu)^{\alpha}\sigma(\varepsilon; \theta, \theta')\label{cond_ene}
\end{eqnarray}

in which, in the linear response theory, the generalized conductivity $\sigma(\varepsilon; \theta, \theta')$ is defined as,
\begin{eqnarray}\label{eq:sigma}
\sigma(\varepsilon; \theta, \theta')&=&e^{2}\int\frac{d^{2}k}{(2\pi)^{2}}\delta\left(\varepsilon-\varepsilon(\vec{k})\right) v^2(\phi)\nonumber\\
&&\quad\left[a(\phi)\cos\theta+b(\phi)\sin\theta\right]\cos(\theta-\xi(\phi))\qquad
\end{eqnarray}
with $\theta=\theta^{\prime}=0$ for $\sigma_{xx}$ and $\theta=\theta^{\prime}=\pi/2$ for $\sigma_{yy}$. We focus here on low enough temperatures, where only electrons contribute effectively in thermal transport and disregard phonon contribution.

\begin{figure}
\includegraphics[width=9.5cm]{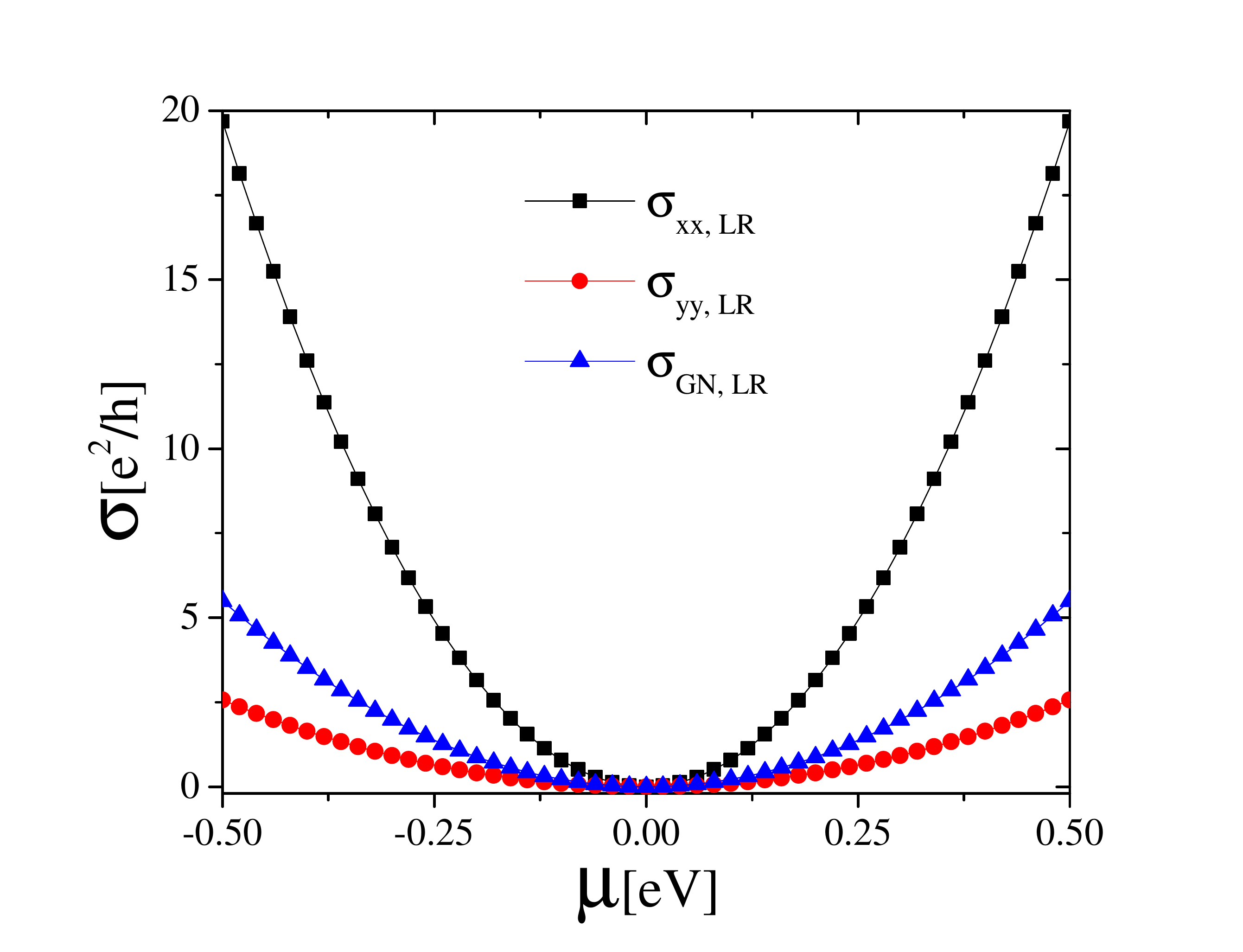}
\caption{(Color online) The conductivity of monolayer borophane as a function of the chemical potential $\mu$, in the presence of long-range impurity potential along the $x$ ($\sigma_{xx}$), and $y$ ($\sigma_{yy}$), directions. The conductivity of the graphene is also shown.}
\label{fig3}
\end{figure}

\begin{figure}
\includegraphics[width=8.5cm]{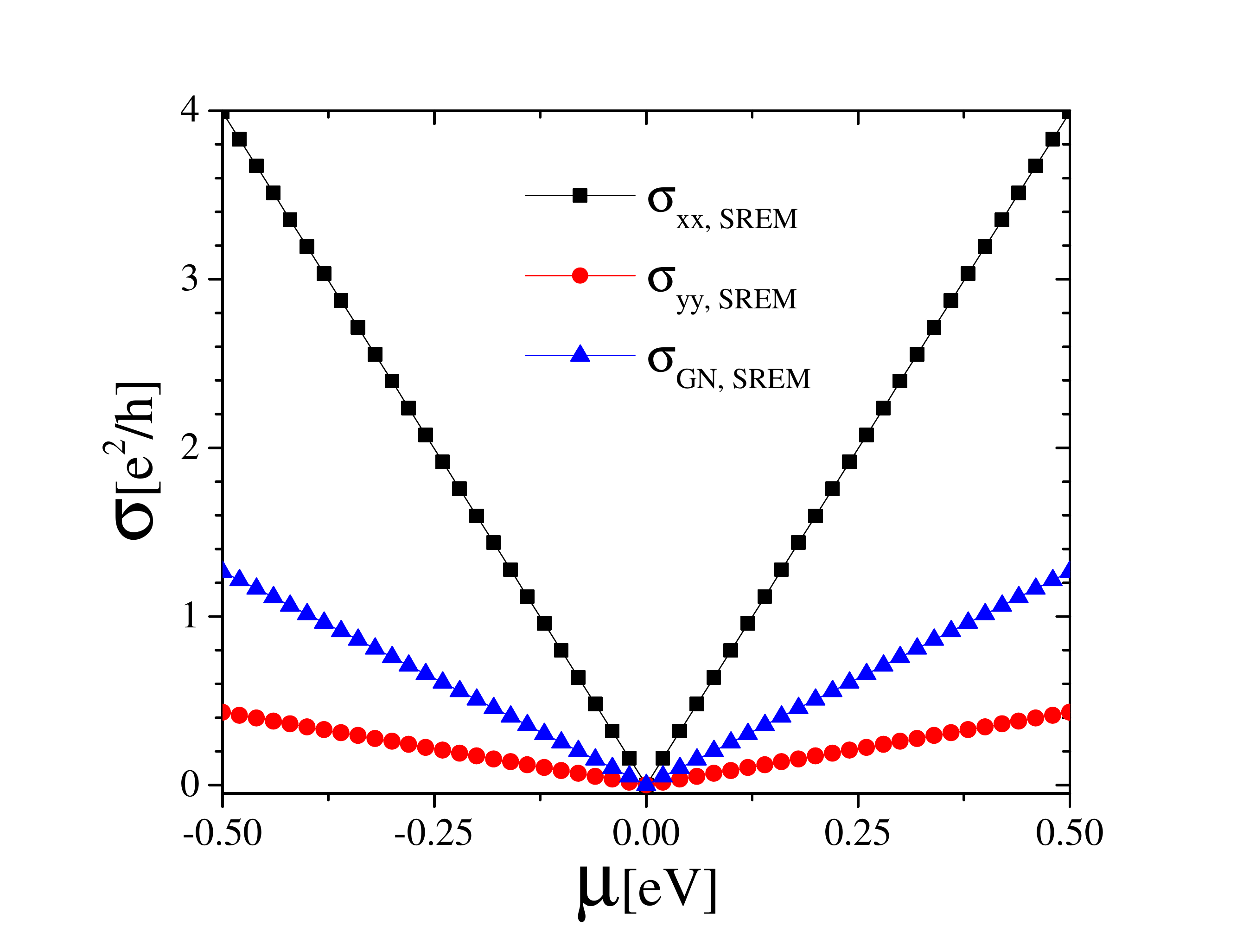}
\caption{(Color online)
The electrical conductivity of borophane along the $x$ ($\sigma_{xx}$), and $y$ ($\sigma_{yy}$), directions as a function of chemical potential $\mu$, in the presence of short-range electro-magnetic scatterers for $\alpha=0.5$. The conductivity of graphene is also shown.}
\label{fig4}
\end{figure}

\subsection{Electro-magnetic scatterers}

In this section, we expand our theoretical analysis to a scattering by magnetic impurities seated on the surface of borophane. Magnetic scattering in dilute charged magnetic impurities, containing short-range electric and ferromagnetically ordered magnetic potentials, whose magnetic moments are along the $i$-th direction, is described using the operator $\hat{V}=V_0 (\alpha + \sigma_i) $ \cite{Rushforth:2007_a,Rushforth:2007_b}, where the dimensionless quantity $\alpha$ is the ratio of the electric and magnetic parts of the impurity potential for an electro-magnetic scatterer.
Considering scattering off a $\delta$-scatterer potential of $\hat{V}$, the matrix elements of $\hat{V}/V_0$ in the basis~(\ref{wave}) can be calculated as follows

\begin{eqnarray*}
\vert\langle {\vec{k}}'|\hat{V}/V_0^x|{\vec{k}}\rangle\big\vert &=&
  \frac{1}{2}\big[\alpha(1+e^{i(\beta_{k'}-\beta_k)})+e^{i\beta_{k'}} + e^{-i\beta_k}\big],\\
\vert\langle {\vec{k}}'|\hat{V}/V_0^y|{\vec{k}}\rangle\big\vert &=&
  \frac{1}{2}\big[\alpha+e^{-i\beta_k}(i+\alpha e^{i\beta_{k'}})-i\cos\beta_{k'}+\sin\beta_k \big],\\
\vert\langle {\vec{k}}'|\hat{V}/V_0^z|{\vec{k}}\rangle\big\vert &=&
  \frac{1}{2}\big[(1+\alpha)e^{i(\beta_{k'}-\beta_k)}+\alpha-1\big],\\
\end{eqnarray*}
It is straightforward to show that the coefficients $ a(\phi), b(\phi)$, corresponding to the electro-magnetic scatterers, in the conductivity formula (Eq.\ref{eq:sigma}), are obtained as
\begin{eqnarray}\label{eq-13}
  a(\phi)&=& \frac{2}{3\pi K}\cos\phi, \nonumber\\
  b(\phi)&=& \frac{2}{\pi K}\sin\phi,
\end{eqnarray}
where the prefactor $K$, can be obtained from the Fermi's golden rule as

\begin{eqnarray}\label{eq-38}
   K = w(\phi,\phi')|\langle \vec{k}'|\hat{V}/V_0| \vec{k}\rangle|^{-2}\,.
\end{eqnarray}
  in which
\begin{eqnarray}\label{eq-38}
  w(\phi,\phi')= \displaystyle
  \frac{1}{(2\pi)^2}\int_0^\infty k'\, dk' w(\vec{k},\vec{k}'),
\end{eqnarray}

Finally from the calculated expressions of $ a(\phi)$ and $b(\phi)$ via Eq.~\ref{eq-13} and using the Boltzmann Eq.~\ref{2dani}, we obtain the following form for the non-equilibrium distribution function;

\begin{equation}\label{eq-14}
  f(\phi,\theta)\!=\!f_0 - ev{\cal E}(-\partial_\epsilon f_0) \frac{2}{\pi K}
  \bigg[\frac{1}{3}\cos\theta\cos\phi + \sin\theta\sin\phi\bigg]\,.
\end{equation}

The Anisotropic Magnetoresistance (AMR) effect is the difference in electrical conductivities, depending on whether the scatterer’s magnetic moments are parallel or anti-parallel to the direction of the current, that is defined as follows

\begin{equation}\label{eq-19}
  \mbox{AMR} = \frac{\sigma_{xx}-\sigma_{yy}}{\sigma_{xx}+\sigma_{yy}}
\end{equation}

Interestingly, for each value of $\alpha$, the AMR is constant and is chemical potential independent for both $n$ and $p$ doped borophane. For example for $\alpha=0.5$, AMR is about $0.805$.

\section{Numerical Results and Discussion}\label{sec:results}

In this section, our numerical results for the thermoelectric transport in borophane are presented. It should be noted that, except for Fig.~\ref{fig9} (that is related to the room temperature figure of merit) we set $T=20 \, K$ in all calculated quantities.
Using the semiclassical Boltzmann approach, we investigate the electrical conductivity, Seebeck coefficient ($\mathcal{S}$), and its corresponding figure of merit ${\it ZT}$, considering the various types of resistive scattering
potential, such as (1) the short-range potential, (2) long-range charge-charge Coulomb potential, with a Thomas- Fermi screening as the source of scattering and (3) impurities containing short-range electric and ferromagnetically ordered magnetic potentials, the electro-magnetic scatterers. It is important to emphasize that this semiclassical theory is valid only in a dilute impurity concentration regime {\it i.e.} when the impurities concentration is much smaller than the charge carriers concentration. Moreover, to ensure that the diluteness criteria is satisfied, the impurities concentration $n_{{\rm imp}}=10^{10}{\rm m^{-2}}$ (corresponding to the chemical potential of approximately $\mu\sim3\times10^{-8}~e{\rm V}$), is considered.

It is worthwhile to mention that the other essential criteria for utilizing the Boltzmann equation are as follows: Particles might interact via binary collisions, impurity density is low in terms of the charge carriers, external fields might have short-range frequencies and all collisions are elastic and involve only uncorrelated particles.
Here, we notice that using the lattice symmetry, the orientations of the two principal lattice axes, which were generally referred to as the armchair ($x$) and zigzag ($y$) axes, are considered.
\par
Due to the fact that the long-range charge-impurity Coulomb interaction is mostly the dominant scatterers in samples, here we consider the Coulomb interaction. To this end, we use an interaction potential, as is commonly used for a 2D electron gas, given by the static Thomas-Fermi (TF) screening type~\cite{ando}
\begin{eqnarray}
V_{{\vec{k}}- {\vec{k}}'}=2\pi e^{2}/[\epsilon(\vert{\vec{k}}-{\vec{k}}'\vert+q_{_{{\rm TF}}})],
\end{eqnarray}
where $q_{_{{\rm TF}}}=2\pi e^{2}D(E)/\epsilon$ is the Thomas-Fermi screening vector with the density of states of the system, $D(E)$.
We assume that the scattering charge centers are at the SiO$_{2}$-borophane interface, thus we use the dielectric constant of this common substrate which is about $\epsilon=2.45$. Meanwhile, we note that we set $T\sim20~{\rm K}$ in all calculated quantities.

In Fig. (\ref{fig3}), the conductivity of borophane (in units of $e^2/h$) as a function of the chemical potential $\mu$, is plotted in the presence of LR potential, along the $x$ ($\sigma_{xx}$), and $y$ ($\sigma_{yy}$), directions. The conductivity of graphene is also shown.
As can be seen, a significant orientation dependent electrical conductivity is observed, where the conductivity in the armchair direction $\sigma_{xx}$ is more than the zigzag conductivity $\sigma_{yy}$. On the other hand, contrary to the electron-hole asymmetry in borophane, its electron-hole conductivity is nearly symmetric. We should mention that at very low temperatures the variation of thermal conductivity will be similar to the charge conductivity $\mathcal{K}\approx (\pi^{2}/3)k_{{\rm B}}T \sigma$ with the Boltzmann constant $k_{{\rm B}}$.

\par
The electrical conductivity of borophane along the zigzag ($\sigma_{yy}$), and armchair ($\sigma_{xx}$), directions versus the chemical potential, in the presence of short-range electro-magnetic scatterers is shown in the figure (\ref{fig4}). In this figure $\alpha=0.5$ and $V_{{\vec{k}}- {\vec{k}}'}=V_{0}=1000~e{\rm V}$\AA$^{2}$~\cite{sarma}. The conductivity of graphene is also shown. As can be seen, the overall energy dependence is the same as LR impurity scattering.

Importantly, the anisotropy ratio of the conductivities ($\sigma_{xx}/\sigma_{yy}$), for both LR and SREM scatters has constant values of $7.67$ and $9.27$, respectively that are independent of chemical potential.
We compare our numerical results with those obtained for monolayer phosphorene ~\cite{Liuprb16}, a highly unique anisotropic 2D material, in the case that $d = 0$, where $d$ is the out-of-plane vertical distance between the charged impurity and monolayer phosphorene. Obtained anisotropic ratio decreases with increasing doping level, varies from 7 to 1.5, depending on the values of $d$ and $n$, the electron (or hole) density. For uniformly distributed impurities in the phosphorene, the anisotropy ratios varies from 3.5 to 4. Meanwhile, our previous work reveals that the anisotropy ratio of the phosphorene conductivities changes from $3$ to $7$ for electrons, in the presence of SR and LR potentials, respectively~\cite{moslem-bp2}, also reported in Refs.~\cite{Luonatcom15,Xianatcom14}.
Thus, in comparison to phosphorene (with a maximum value of 7 for the anisotropy ratio of conductivities ), borophane with a high anisotropy ratio of about $10$, is an unprecedented anisotropic material.

To the best of our knowledge, there is still no experimental probe about the thermoelectric effects in borophane for now. But there are some theoretical studies exploring the electrical and thermal conductivities of borophane~\cite{Sun18,Padilhapccp16,T.Cheng17} . Our numerical calculations have been faithfully confirmed by several research groups.

Based on the first-principles calculations, and according to the I-V curves, J. Sun {\it et al.}~\cite{Sun18} argued that the electrical conductivity in the armchair direction is greater than that of the zigzag direction ($\sigma_{xx}>\sigma_{yy}$).
This directional dependency of the electric conductivity of borophane also reported in Ref.~\cite{Padilhapccp16}.
Our findings is fully consistent, both quantitatively and qualitatively, with those obtained in Refs.~\cite{Sun18,Padilhapccp16}.

Let us here, compare our results for the short- range and long-range scatterers in borophane with those in graphene.
In a comparative study of the conductivity of graphene, in the tight-binding Landauer approach and on the basis of the Boltzmann equation, it has been argued by J. W. Klos, { \it et al.}~\cite{Klos} that in the case of short-range scattering, the electrical conductivity of graphene is independent of the chemical potential. It is worth to emphasize that for the short-range scatters, just like graphene, the short-range conductivities of borophane are the constant values (independent of the chemical potential),that are $\sigma_{xx}=250\ e^2/h$ and $\sigma_{yy}=33\ e^2/h$, along the armchair and zigzag direction, respectively at $T\sim20={\rm K}$.
In the case of long-range conductivity, our finding also is in a good agreement with the majority of experimental findings~\cite{SarmaRev}.

\par
\begin{figure}
\includegraphics[width=9.5cm]{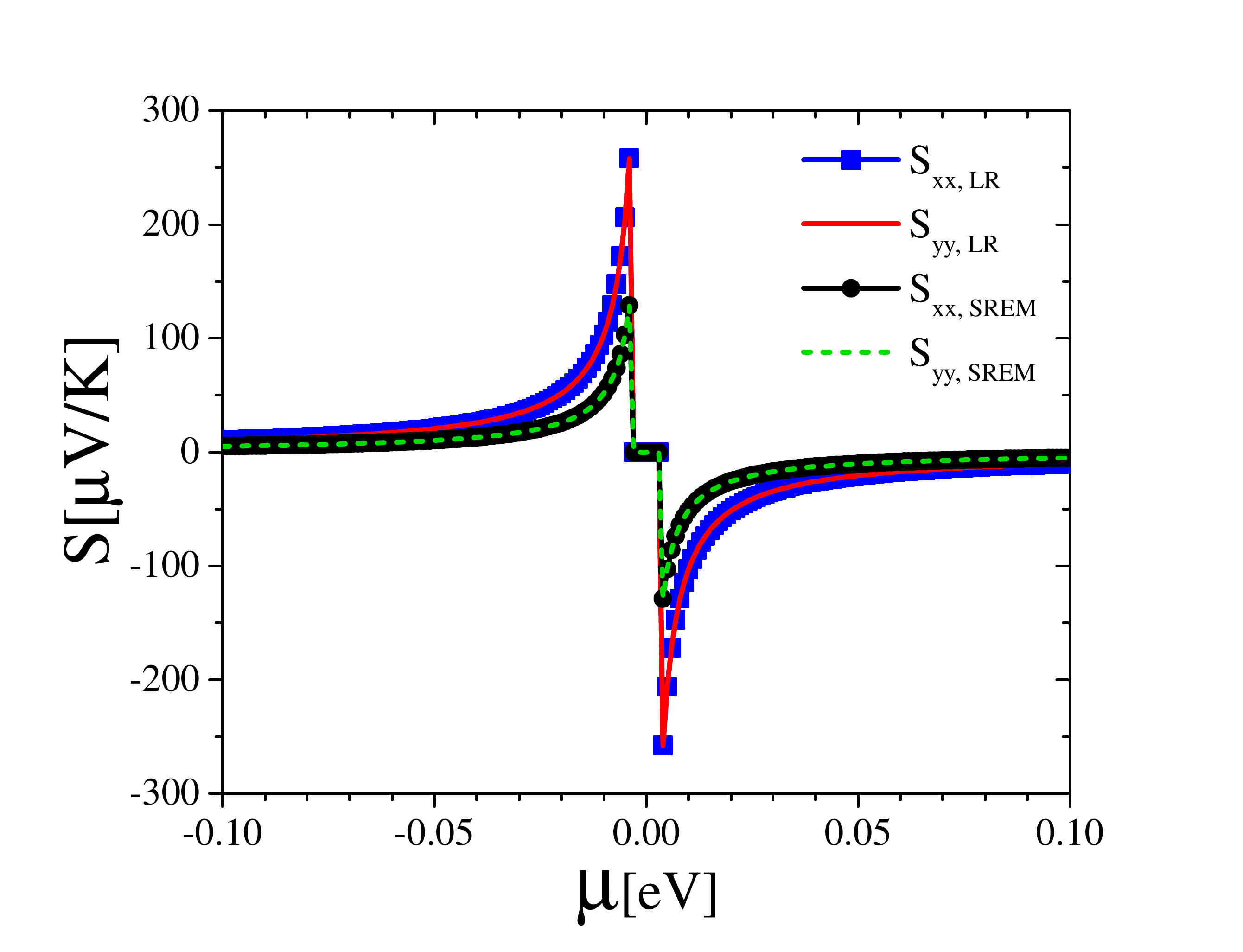}
\caption{(Color online) Seebeck coefficient of borophane as a function of chemical potential $\mu$, in the presence of short-
and long-range Coulomb potentials. Despite the type of the impurity scatterers, the Seebeck coefficients are completely isotropic for both directions.}
\label{fig7}
\end{figure}

As a more feasible quantity in the real experiments, the variation of the Seebeck coefficients $\mathcal{S}$, with chemical potential $\mu$, in the presence of LR and SREM potentials is obtained, shown in Fig. (\ref{fig7}). In convention, the sign of the Seebeck coefficient is the sign of the potential of the cold end with respect to the hot end of the temperature gradient. Thus a negative charge thermopower is obtained for when the Fermi energy lies in the conduction band because thermally activated electrons move along the temperature gradient which results in a charge accumulation gradient in the opposite direction due to the negative charge, however, thermally activated holes in the valence band lead to a positive thermopower.
An important feature of the borophane thermopower is its isotropic behavior of three times lower in the SREM scatterers compared to the LR scatterers.

\begin{figure}
\includegraphics[width=9.5cm]{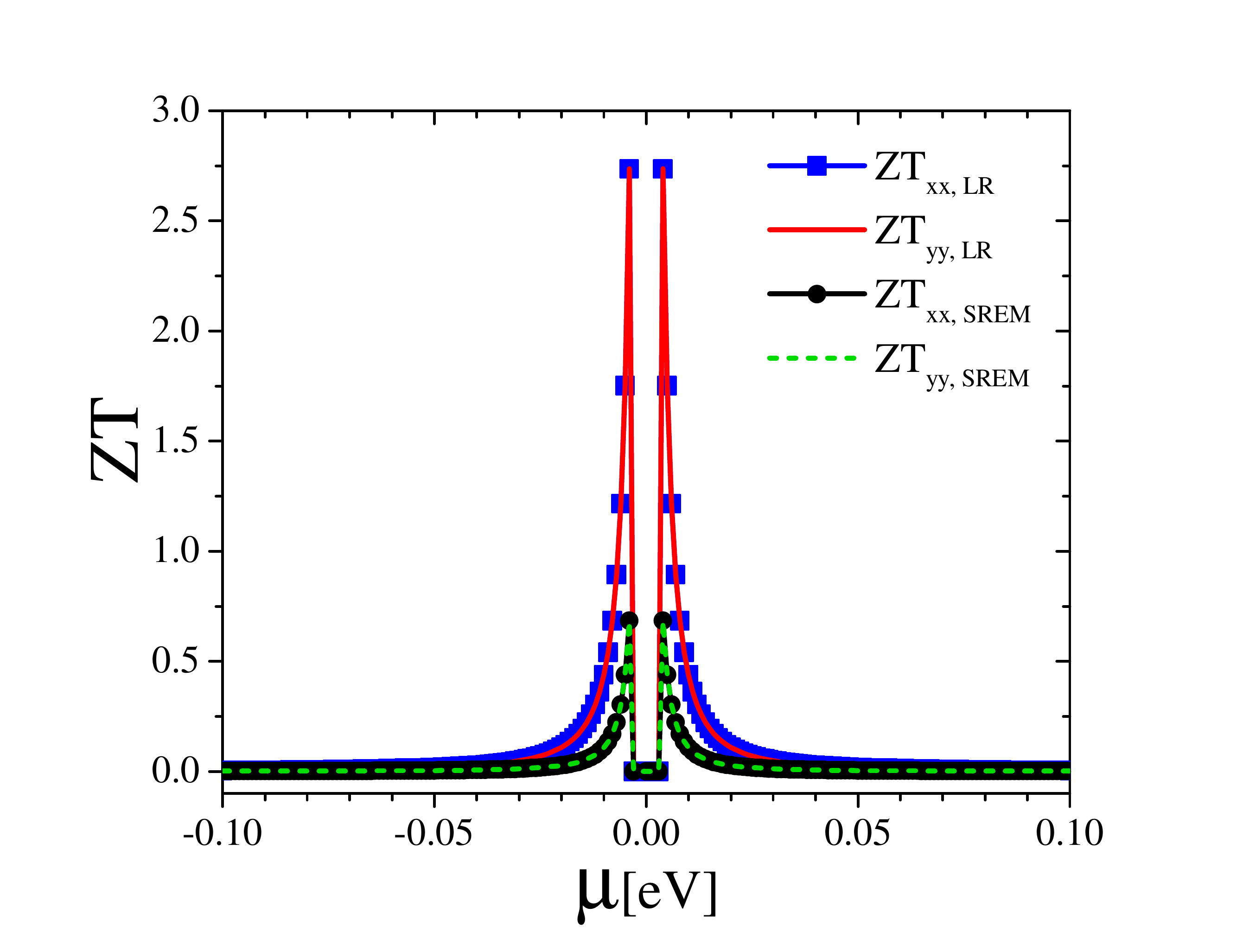}
\caption{(Color online) The variation of corresponding figures of merit as a function of chemical potential $\mu$, in the presence of long-range charge impurity and electro-magnetic scatterers. Figures of merit are also nearly isotropic.}
\label{fig8}
\end{figure}

\begin{figure}
\includegraphics[width=9.5cm]{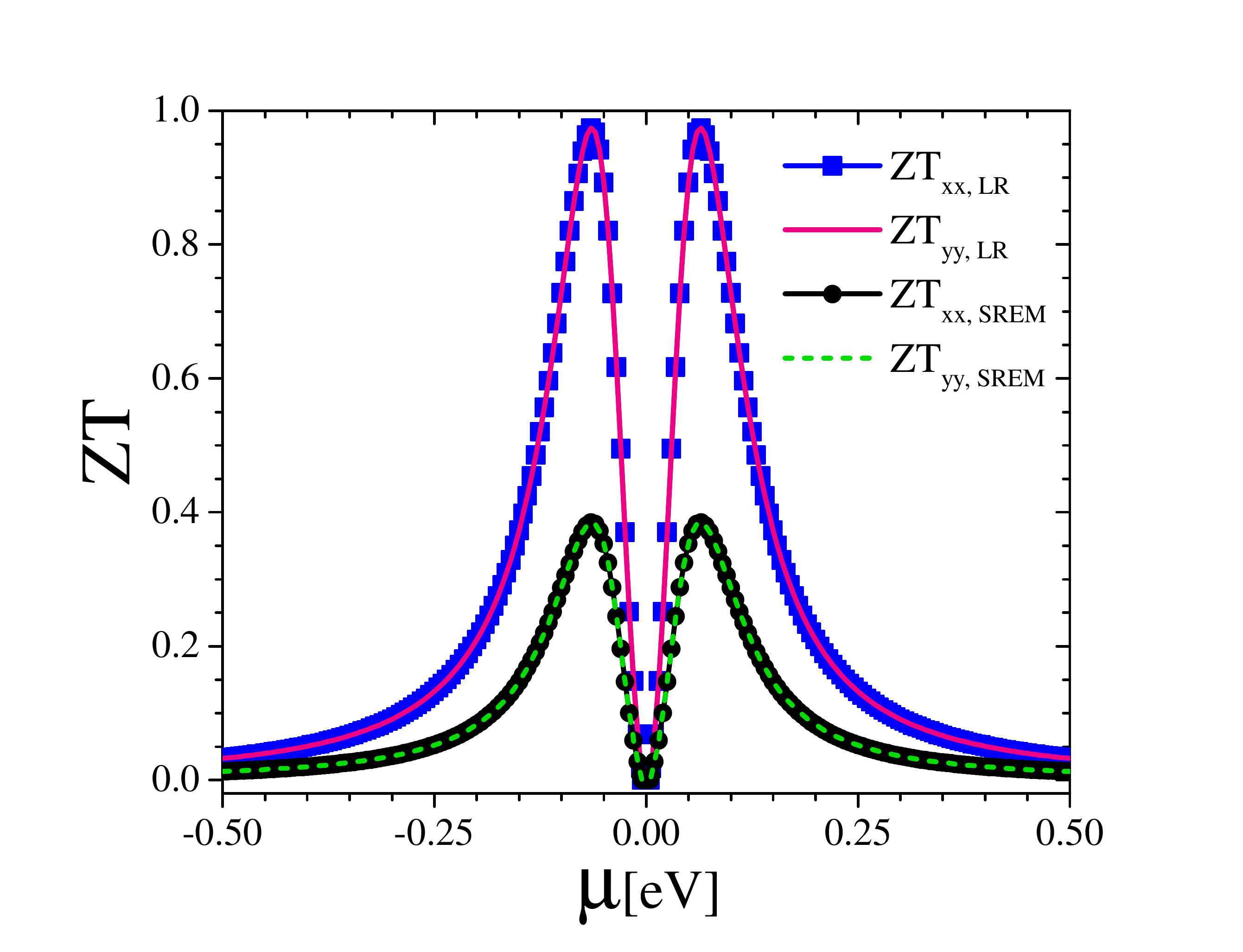}
\caption{(Color online) The variation of room temperature figures of merit along the armchair and zigzag directions as a function of chemical potential $\mu$, in the presence of long-range and short-range electro-magnetic impurity potential. The effect of both electron and phonon contribution in the thermal conductivity ($\mathcal{K}_{{\rm tot}}=\mathcal{K}_{{\rm el}}+\mathcal{K}_{{\rm ph}}$) is considered. Phonon contribution of the thermal conductivity is $\sim 150$ and $\sim 267$ Wm$^{-1}$K$^{-1}$ along the armchair and zigzag directions, respectively.}
\label{fig9}
\end{figure}

Moreover, the figure of merit ${\it ZT}$ is depicted as a function of the chemical potential $\mu$, for both LR and SREM scatterers in Fig.\ref{fig8}. Our findings reveal that, in contrast to highly anisotropic electrical and thermal conductivities, the Seebeck coefficient and its corresponding thermoelectric figures of merit are completely isotropic, consistent with the isotropic thermopower and figures of merit in phosphorene, which has a highly anisotropic conductivity~\cite{Faghaninia,moslem-bp2,moslem-stt}. Interestingly, the figure of merit attains its maximum value around the charge neutral point as can be seen in Fig.\ref{fig8}.

\subsection{ The phonon contribution to the intrinsic resistivity and figure of merit }\label{sec:concl}
Subsequently, we look into the contributions of phonon to the intrinsic resistivity of borophane. For a better understanding, we review the phonon-limited resistivity in metals, graphene and borophene and other 2D materials.
According to the kinetic theory for semiconductors and insulators, heat is mainly carried by phonons- the propagating local distortions of the crystal ~\cite{Ziman}.

In the metals at room temperature, the scattering of electrons by phonons is typically the dominant source of resistivity. At temperatures greater than the Debye temperature $\theta_D$, upon which all phonons are excited to scatter carriers, the electrical resistivity is proportional to temperature and for the temperatures below the Debye temperature, the phonon modes begin to freeze out and the resistivity drops much more rapidly whereas for a three-dimensional metal is expected to drop as $\rho (T)\propto T^5$ and for a two-dimensional metal varies as $\rho (T)	\propto T^4$, the so-called Bloch-Grüneisen regime ~\cite{Bloch30,Gruneisen}.

For a semimetal with a small cylindrical Fermi surface, at intermediate $T$, there is a temperature window in which the resistivity varies purely as $\rho (T)	\propto T^2$, arises from the anisotropy of the Fermi-surface.
As reported in Ref.~\cite{Zhang-Angew18}, only the electron/hole pocket at the center of the Brillouin zone is mainly responsible for the intrinsic transport properties.
Although the electron pockets in borophane are ellipsoids, the same physics that yields the unusual resistivity for a cylindrical Fermi surface also applies to an ellipsoid~\cite{Kukkonen49}.

Previous theoretical and experimental studies ~\cite{Hwang08prl,Efetov10}, show that the intrinsic electrical resistivity of graphene, arising from electron-phonon interactions, is proportional to $\rho (T)	\propto T^4$ at low temperatures (reflecting the 2D nature of the electrons and the acoustic phonons in graphene), while varies linearly with $T$ at high temperatures $\rho (T) \propto T$, (a semiclassical behavior), that is independent of doping. It is worth mentioning that, graphene has a Debye temperature $\theta_D=$2300 K, approximately an order of magnitude higher than for typical metallic materials.
The Debye temperature of borophene have reported as 863.86 K and 2000 K ~\cite{H.Zhounpj,H.Zhounpj,Toheiphon}, which is much higher than that of monolayer $MoS_2$ (262.3 K)~\cite{PengRSC16} and black phosphorene (500 K)~\cite{JainSci15}, but lower than that of graphene.
It is firmly believed that~\cite{Zhang-Angew18} for all the 2D boron allotropes, the Debye temperature is about $\theta_D=$1700 K which corresponds to the highest phonon energy $\sim$1200 $cm^{-1} $.

A quadratic dispersion acoustic branch is always present on suspended few-layer systems, having a strong impact on the calculated thermal conductivity of some 2D systems.
The theoretical and experimental observations reveal that the thermal conductivity of 2D materials is highly sensitive to the crystal structure in addition to the physical factors including the atomic masses, width dependence, edge chirality, roughness, hydrogen passivation etc, giving rise to a large gap between the theoretically obtained results and the experimental data.

To explore the thermal conductivity of infinitely large 2D crystals through the Peierls-Boltzmann transport equation (PBTE), rigorously one needs to employ either the iterative approach ~\cite{Omini,Broido} or a variational method ~\cite{Fugallo13,Fugallo14}.
Therefore, the study of thermal conductivity of borophane has been limited to just numerical simulations.

The small atomic mass of Boron atoms in borophene causes a strong electron-phonon coupling.
Different from graphene, the absolute value of the phonon resistivity ($\rho_{e-ph}$) of borophene is highly sensitive to external carrier densities, because  an ultrahigh doping level in graphene leads to a negligible variation in $\rho_{e-ph}$. Furthermore, the resistivity is highly dependent on the different two-Dimensional boron Polymorphs and the $\rho_{e-ph}$ of various polymorphs of borophene can be greatly varied by carrier concentration. To our knowledge, experimental or theoretical investigations on the phononic transport properties of borophene are still lacking.

It would be also worthwhile to mention that due to the strong bonding in borophene and borophane, Debye temperature in these systems has a high value about 2000 K ~\cite{H.Zhounpj,Toheiphon}. Therefore, the room temperature (300K) can be considered as low with respect to high Debye temperature in boron allotropes, and therefore, lattice vibrations should be treated quantum-mechanically. As a result, the phonon population is low at room temperature, due to the high Debye temperature of borophane, which suppresses the possibility of phonon-phonon inelastic scattering events. Furthermore, as material or device dimensions decrease, in particular, when the length scale is comparable or smaller than the phonon mean free path, the inelastic scattering will be further suppressed. Thus, the thermal transport is quasi-ballistic or ballistic, which cannot be explained by phonon diffusion theory.

Moreover, the study of these materials necessarily requires the incorporation of quantum effects into thermal transport analysis.
Therefore, it is difficult to directly estimate the behavior of the electrical conductivity and thus ZT of borophane with temperature.
The phonon contribution of the thermal conductivity can only affect the figure of merit, and the Seebeck coefficient and power factor $\mathcal{S}^2\sigma$ is not altered by the presence of the phonons~\cite{moslem-bp2}. At high temperatures, the phonon becomes important but it only results in the overall decline of the figures of merit, without affecting their qualitative behavior. Due to the fact that the thermal conductivity is approximately independent of doping~\cite{moslem-bp2,moslem-physicac}, by using the room temperature contribution of the phonons in the thermal conductivity of borophane with ($\mathcal{K}_{{\rm ph}}\sim$ $150$ and $\sim 267$ Wm $^{-1}$K$^{-1}$ along the armchair and zigzag directions, respectively~\cite{Wangpccp16}), we estimate the figure of merit of borophane at room temperature.

In Fig. (\ref{fig9}), the room temperature ($T=300$~K) figure of merit versus the chemical potential $\mu$, is plotted in the presence of LR and SRMG scatterers, along the zigzag (${\it ZT}_{yy}$) and armchair (${\it ZT}_{xx}$) directions. Notice also that the electronic contribution of the thermal conductivity is obtained numerically from Eq. (\ref{eq:L}).
It can be seen that just like the low temperature figure of merit, the room temperature figure of merit is completely isotropic. Compared with graphene, a high value ${\it ZT}$ of about $1$ can be achieved in a monolayer borophane at room temperature.

As already pointed out, semi-metals with large electron-hole asymmetry can be considered as an important strategy for strong enhancement of the thermoelectric
coefficients~\cite{Markov-scr18}. Similarly, borophane as a semi-metals with the large electron-hole asymmetry, attains high value of ${\it ZT}$ at room temperature (${\it ZT}\sim 1$).

\section{Conclusion}\label{sec:concl}

In conclusion, we have investigated the impact of impurity scattering on charge carrier transport in an anisotropic Dirac system, using the generalized Boltzmann approach. As a case study, we have investigated the thermoelectric performance of borophane, a new monolayer Dirac semimetal with two tilted and anisotropic Dirac cones, in the diffusive transport regime. At low temperature elastic scattering becomes the dominant mechanism as inelastic scattering is strongly suppressed. In this work our focus is on elastic scattering and therefore, with good approximation, intervalley scattering (interband processes) is neglected. Electron scattering from the various impurities located at the surface of a monolayer borophane, is different for massless fermions with tilted Dirac cone.
Finding the exact solution to the linear-response Boltzmann equation, the electrical conductivity and thermoelectric properties of borophane in the presence of the short-range, long-range charged impurity and the short-range electro-magnetic scatterers were studied.
Contrary to the electron-hole asymmetry in borophane, its electron-hole conductivity is nearly symmetric. Interestingly, for the short-range scatters, just like graphene, the short-range conductivities of borophane are the constant values and does not depend on the chemical potential. The conductivities of the SREM scatterers have a linear dependence on the doping levels.

We explained the effect of the chemical potential on the thermoelectric properties of borophane. We demonstrated that, regardless of the impurity type, the electric conductivity of borophane was highly anisotropic, while the Seebeck coefficient and figure of merit (${\it ZT}$) were nearly isotropic.
The anisotropy ratio of the conductivities ($\sigma_{xx}/\sigma_{yy}$) for both the long-range and short-range magnetic impurities were constant values of around $7.67$ and $9.27$, respectively.
Along with the ambipolar nature of the borophane thermopower, it was found to have a large thermoelectric figure of merit of about ${\it ZT}=2.75$ and ${\it ZT}=1$, at low temperature ($T\sim20~{\rm K}$) and room temperature, respectively, due to the large asymmetry between electrons and holes in borophane.
More importantly, borophane attained its maximum value of ${\it ZT}$ at very low chemical potentials, in the vicinity of the charge neutrality point.
In comparison to phosphorene (a highly unique anisotropic 2D material) borophane with a high anisotropy ratio of about $10$, is an unprecedented anisotropic material. This high anisotropy ratio together with the large figure of merit, suggest that borophane is promising for the thermoelectric applications and transport switching in the Dirac transport channels.
The results found are in good agreement with recent theoretical and experimental data on borophane samples, and elucidate the role of the different phonon modes in limiting electron mobility.

\section{acknowledgments}
This work was partially supported by Iran Science Elites Federation.


\begin{thebibliography}{99}
\providecommand \Doi[1]{\href{\doibase#1}}

\bibitem{Gonzalez2007prl}
 N. Gonzalez Szwacki, A. Sadrzadeh, B. I. Yakobson, \Doi{10.1103/PhysRevLett.98.166804}{Physical Review Letters, 98, (16) 166804 (2007).}

\bibitem{Gonzalez2007nrl}
 N. Gonzalez Szwacki, Nanoscale Research Letters, \Doi{10.1007/s11671-007-9113-1}{ {\bf 3}, 49 (2007).}

\bibitem{mannix2015synthesis}
A. J. Mannix, X-F Zhou, B. Kiraly, J. D. Wood, D. Alducin, B. D. Myers, X. Liu, B. L. Fisher, U.
Santiago, J. R. Guest, M. J. Yacaman, A. Ponce, A. R. Oganov, M. C. Hersam, and N. P. Guisinger,\Doi{10.1126/science.aad1080}{Science {\bf 350}, 1513 (2015).}

\bibitem{Otten2002}
C. J. Otten, O. R. Lourie, M. F. Yu, J. M. Cowley, M. J. Dyer, R. S. Ruoff, and W. E. Buhro,\Doi{10.1021/ja017817s}{ J. Am. Chem. Soc. {\bf 124} (17), pp 4564–4565 (2002).}

\bibitem{F.Liu-nanowire}
F. Liu, D. M. Tang, H. Gan, X. Mo, J. Chen, S. Deng†, N. Xu, Y. Bando, and D. Golberg, \Doi{10.1021/nn404316a}{ACS Nano, {\bf 7} (11), pp 10112 (2013)}.

\bibitem{Ciuparu2004}
D. Ciuparu, R. F. Klie,  Y. Zhu,  L. Pfefferle, \Doi{10.1021/jp049301b}{J. Phys. Chem. B {\bf 108}(13), 3967 (2004).}

\bibitem{F.Liu2010}
F. Liu, C. Shen,Z. Su, X. Ding, S. Deng, J. Chen, N. Xu, H. Gao, \Doi{10.1039/B919260C}{J. Mater. Chem. {\bf 20}(6267), pp 2197 (2010).}

\bibitem{Kiran2005}
B. Kiran, S. Bulusu, H. J. Zhai, S. Yoo, X. C. Zeng, L. S. Wang,  \Doi{10.1073/pnas.0408132102}{Proc. Natl. Acad. Sci. U.S.A. {\bf 4}, 961 (2005).}

\bibitem{Oger2007}
E. Oger, N. R. M. Crawford, P. Kelting, P. Weis, M. M. Kappes, R. Ahlrichs, \Doi{10.1002/anie.200701915}{ Angew. Chem., Int. Ed.{\bf 46}, 8503 (2007).}

\bibitem{An2006}
W. An, S. Bulusu, Y. Gao, X. C. Zeng, \Doi{10.1063/1.2187003}{ J. Chem. Phys. {\bf 124}, 154310 (2006).}
\bibitem{Eremets2001}
 M. I. Eremets, V. V. Struzhkin, H.-k. Mao, R. J. Hemley,  \Doi{10.1126/science.1062286}{ Science {\bf 293}(5528), 272 (2001).}

\bibitem{H.Liu13}
H. Liu, J. Gao, \Doi{10.1038/srep03238}{J. Zhao, Sci. Rep. {\bf  3}, 3238 (2013).}

\bibitem{X.Wu12}
X. Wu, J. Dai, Y. Zhao, Z. Zhuo, J. Yang, X. Zeng, \Doi{10.1021/nn302696v}{ACS Nano {\bf 6}, 7443 (2012).}

\bibitem{Y.Liu13}
Y. Liu, E. S. Penev, B. I. Yakobson,\Doi{10.1002/anie.201207972}{ Angew. Chem., Int. Ed. {\bf 52}, 3156 (2013).}

\bibitem{Z.Zhang15}
Z. Zhang, Y. Yang, G. Gao, B. I. Yakobson, \Doi{10.1002/anie.201505425}{Angew. Chem., Int. Ed.  {\bf 54}, 13022 (2015).}

\bibitem{Penev-NL16}
 E. S. Penev,  A. Kutana, B. I. Yakobson,\Doi{10.1021/acs.nanolett.6b00070}{ Nano Lett. {\bf 16}, 2522 (2016).}

\bibitem{M.Gao}
M. Gao, Q.-Z. Li, X.-W. Yan, J. Wang, \Doi{10.1103/PhysRevB.95.024505}{ Phys. Rev. B {\bf 95}, 024505 (2017).}

\bibitem{RC.Xiao}
R. C. Xiao, D. F. Shao, W. J. Lu, H. Y. Lv, J. Y. Li, Y. P. Sun,\Doi{10.1063/1.4963179}{ Appl. Phys. Lett. {\bf 109}, 122604 (2016).}

\bibitem{ZZhang}
Z. Zhang, E. S. Penev, and B. I. Yakobson,\Doi{10.1038/nchem.2521}{Nat. Chem. {\bf 8}, 525 (2016).}

\bibitem{BFeng}
B. Feng, J. Zhang, Q. Zhong, W. Li, S. Li, H. Li, P. Cheng, S. Meng, L. Chen, and K. Wu, \Doi{10.1038/nchem.2491}{Nat. Chem. {\bf 8}, 563 (2016).}

\bibitem{BFeng2} B. Feng, J. Zhang, R.-Y. Liu, T. Iimori, C. Lian, H. Li, L.
Chen, K. Wu, S. Meng, F. Komori, and I. Matsuda, \Doi{10.1103/PhysRevB.94.041408}{Phys. Rev. B {\bf 94}, 041408(R) (2016).}

\bibitem{Boro}
B. Feng, O. Sugino, R. Y. Liu, J. Zhang, R. Yukawa, M. Kawamura, T. Iimori, H. Kim, Y. Hasegawa, H. Li, L. Chen, K. Wu,
H. Kumigashira, F. Komori, T. C. Chiang, S. Meng, I. Matsuda., \Doi{10.1103/PhysRevLett.118.096401}{Phys. Rev. Lett. {\bf 118}, 096401 (2017).}

\bibitem{Boustani1}
I. Boustani., \Doi{10.1103/PhysRevB.55.16426}{Phys. Rev. B {\bf 55}(24), 16426 (1997).}

\bibitem{Boustani2}
I. Boustani., \Doi{10.1016/S0039-6028(96)00969-7}{Surf. Sci. {\bf 370}(2–3), 355 (1997).}

\bibitem{Piazza14}
Z. A. Piazza, H. S. Hu, W. L. Li, Y. F. Zhao, J. Li, and L. S. Wang, \Doi{10.1038/ncomms4113}{Nat. Commun. {\bf 5}, 3113 (2014).}

\bibitem{Tang2007}
H. Tang and S. Ismail-Beigi., \Doi{10.1103/PhysRevLett.99.115501}{Phys. Rev. Lett. {\bf 99}(11),115501 (2007).}

\bibitem{Tang2009}
H. Tang and S. Ismail-Beigi., \Doi{10.1103/PhysRevB.80.134113}{Phys. Rev. B {\bf 80}(13), 134113 (2009).}

\bibitem{Tang2010}
H. Tang and S. Ismail-Beigi., \Doi{10.1103/PhysRevB.82.115412}{Phys. Rev. B {\bf 82}(11), 115412 (2010).}

\bibitem{X.Wu2012}
X. Wu, J. Dai, Y. Zhao, Z. Zhuo, J. Yang, and X. C. Zeng,  \Doi{10.1021/nn302696v}{ACS Nano{\bf  6}(8), 7443 (2012).}

\bibitem{zhou2014semimetallic}
X. Zhou, X. Dong, A. Oganov, Q. Zhu, Y. Tian, and H. Wang, \Doi{10.1103/PhysRevLett.112.085502}{Phys. Rev. Lett. {\bf 112}, 085502 (2014).}

\bibitem{Sadrzadeh-NL12}
E. S. Penev, S. Bhowmick, A. Sadrzadeh, B. I. Yakobson, Polymorphism of two-dimensional boron, \Doi{10.1021/nl3004754}{Nano Lett. {\bf 12} (5) 2441 (2012).}

\bibitem{Sadhukhan2017}
K. Sadhukhan and A. Agarwal, \Doi{10.1103/PhysRevB.96.035410}{Phys. Rev. B {\bf 96}, 035410 (2017).}

\bibitem{Xu16}
L. C. Xu, A. Du, and L. Kou, \Doi{10.1039/C6CP05405F}{Phys. Chem. Chem. Phys. {\bf 18}, 27284 (2016).}

\bibitem{Jena2017}
N. K. Jena, R. B. Araujo, V. Shukla and R. Ahuja, \Doi{10.1021/acsami.7b01421}{ACS Appl. Mater. Interfaces, {\bf 9}, 16148 (2017).}

\bibitem{Anpccp2018}
Y. An and J. Jiao, Y. Hou, H. Wang, D. Wu, T. Wang, Z. Fu, G. Xu, R. Wu., \Doi{10.1039/C8CP04272A}{Phys. Chem. Chem. Phys. {\bf 20}, 21552 (2018).}

\bibitem{WangSci16}
Z. Wang, T.-Y. Lü, H.-Q. Wang, Y. P. Feng  and J.-C. Zheng,\Doi{10.1038/s41598-017-00667-x}{ Sci. Rep. {\bf 7}, 609 (2016).}

\bibitem{WangRSC17}
Z. Wang, T.-Y. Lü, H.-Q. Wang, Y. P. Feng  and J.-C. Zheng,\Doi{10.1039/C7RA05704K}{ RSC Adv. {\bf 7}, 47746  (2017).}

\bibitem{Dresselhaus93}
L. D. Hicks and M. S. Dresselhaus, \Doi{10.1103/PhysRevB.47.12727}{Phys. Rev. B {\bf 47}, 12727 (1993)}.

\bibitem{Venkatasubramanian}
R. Venkatasubramanian, E. Siivola, T. Colpitts, and B. O'Quinn, \Doi{10.1038/35098012}{Nature (London) {\bf 413}, 597 (2001)} .

\bibitem{Arita}
R. Arita, K. Kuroki, K. Held, A. V. Lukoyanov, S. Skornyakov, and V. I. Anisimov, \Doi{10.1103/PhysRevB.78.115121}{Phys. Rev. B {\bf 78}, 115121 (2008)}.

\bibitem{Hamada}
N. Hamada, T. Imai, and H. Funashima, \Doi{10.1088/0953-8984/19/36/365221}{J. Phys.: Condens. Matter {\bf 19}, 365221 (2007)}.

\bibitem{Zide}
J. M. O. Zide, D. Vashaee, Z. X. Bian, G. Zeng, J. E. Bowers, A. Shakouri, and A. C. Gossard, \Doi{10.1103/PhysRevB.74.205335}{Phys. Rev. B {\bf 74}, 205335 (2006)}.

\bibitem{Wei}
P. Wei, W. Bao, Y. Pu, C. N. Lau, and J. Shi, \Doi{10.1103/PhysRevLett.102.166808}{Phys. Rev. Lett. {\bf 102}, 166808 (2009)}.

\bibitem{Zuev}
Y. M. Zuev, W. Chang, and P. Kim, \Doi{10.1103/PhysRevLett.102.096807}{Phys. Rev. Lett. {\bf 102}, 096807 (2009)}.

\bibitem{Kato}
T. Kato, S. Usui, and T. Yamamoto, \Doi{10.7567/JJAP.52.06GD05}{Jpn. J. Appl. Phys. {\bf 52}, 06GD05 (2013)} .

\bibitem{Buscema}
M. Buscema, M. Barkelid, V. Zwiller, H. S. J. van der Zant, G. A. Steele, and A. Castellanos-Gomez, \Doi{10.1021/nl303321g}{Nano Lett. {\bf 13}, 358 (2013)}.

\bibitem{Konabe1}
S. Konabe and T. Yamamoto, \Doi{10.1103/PhysRevB.90.075430}{Phys. Rev. B {\bf 90}, 075430 (2014)}.

\bibitem{Harman}
L. D. Hicks, T. C. Harman, X. Sun,  M. S. Dresselhaus, \Doi{10.1103/PhysRevB.53.R10493}{Phys. Rev. B {\bf 53}, 10493(R) (1996)}.

\bibitem{Dresselhausprb47-93-1}
L. D. Hicks and M. S. Dresselhaus, \Doi{10.1103/PhysRevB.47.12727}{Phys. Rev. B {\bf 47}, 12727 (1993)}.

\bibitem{Bilc}
D. Bilc, S. D. Mahanti, K. F. Hsu, E. Quarez, R. Pcionek, and M. G. Kanatzidis, \Doi{10.1103/PhysRevLett.93.146403}{Phys. Rev. Lett. {\bf 93}, 146403 (2004)}.

\bibitem{Mahan}
G. D. Mahan and J. O. Sofo, \Doi{http://www.pnas.org/content/93/15/7436.abstract}{Proc. Nat. Acad. Sci. U.S.A. {\bf 93}, 7436 (1996)}.

\bibitem{Fomin}
V. M. Fomin and P. Kratzer, \Doi{10.1103/PhysRevB.82.045318}{Phys. Rev. B {\bf 82}, 045318 (2010)}.

\bibitem{Zianni}
X. Zianni, \Doi{10.1063/1.3523360}{Appl. Phys. Lett. {\bf 97}, 233106 (2010)}.

\bibitem{Neophytou}
N. Neophytou and H. Kosina, \Doi{10.1103/PhysRevB.83.245305}{Phys. Rev. B {\bf 83}, 245305 (2011)}; \Doi{10.1063/1.4737122}{J. Appl. Phys. {\bf 112}, 024305 (2012)}

 \bibitem{Heremanssci08}
J. P. Heremans, V. Jovovic, E. S. Toberer, A. Saramat, K. Kurosaki, A. Charoenphakdee, S. Yamanaka, G. J. Snyder.\Doi{}{ Science {\bf 321}, 554 (2008).}

\bibitem{Liu-Zhao-APL08}
W.-S. Liu, L.-D. Zhao, B.-P. Zhang, H.-L. Zhang, J.-F. Li\Doi{}{ Appl. Phys. Lett. {\bf 93}, 042109 (2008).}

\bibitem{Dresselhaus07}
M. S. Dresselhaus, G. Chen, M. Y. Tang, R. Yang, H. Lee, D. Wang, Z. Ren, J.-P. Fleurial, and P. Gogna,\Doi{}{ Adv. Mater. {\bf 19}, 1043 (2007).}

\bibitem{Dresselhausprb47-93-2}
L. D. Hicks, and  M. S. Dresselhaus,\Doi{10.1103/PhysRevB.47.16631}{ Phys. Rev. B {\bf 47}, 16631 (1993).}

\bibitem{Xiao-npj18}
Y. Xiao and L.-D. Zhao,\Doi{10.1038/s41535-018-0127-y}{ npj. Quan. Mate. {\bf 3}, 55 (2018).}

\bibitem{LaLonde}
A. D. LaLonde, Y. Pei, H. Wang, G. J. Snyder,\Doi{10.1016/S1369-7021(11)70278-4}{ Mater. Today {\bf 14}, 526 (2011).}

\bibitem{Markov-scr18}.
M. Markov, X. Hu, H.-C. Liu, N. Liu, S. J. Poon, K. Esfarjani and M. Zebarjadi,\Doi{10.1038/s41598-018-28043-3}{ Sci. Rep. {\bf 8}, 9876 (2018).}

\bibitem{E.H.Hwang08}
E. H. Hwang and S. Das Sarma, \Doi{10.1103/PhysRevB.77.195412}{Phys. Rev. B 77, 195412 (2008). }

\bibitem{Padilhapccp16}
J. E. Padilha, R. H. Miwa, and A. Fazzio, \Doi{10.1039/C6CP05092A}{Phys. Chem. Chem. Phys. 18, 25491 (2016) }

\bibitem{zabolotskiy2016strain}
A. Zabolotskiy and Y. Lozovik, \Doi{10.1103/PhysRevB.94.165403}{Phys. Rev. B {\bf 94}, 165403 (2016).}

\bibitem{Peeters-Hamilton}
M. Nakhaee, S. A. Ketabi,  and F. M. Peeters, \Doi{10.1103/PhysRevB.97.125424}{Phys. Rev. B {\bf 97}, 125424 (2018).}

\bibitem{Ezawa2017}
M. Ezawa,  \Doi{10.1103/PhysRevB.96.035425}{Phys. Rev. B {\bf 96}, 035425 (2017).}

\bibitem{Islam2017}
S. K. F. Islam and A. M. Jayannavar, \Doi{10.1103/PhysRevB.96.235405}{Phys. Rev. B {\bf 96}, 235405 (2017).}

\bibitem{X.-F.Zhou2016}
X.-F. Zhou and H.-T.Wang, \Doi{10.1080/23746149.2016.1209432}{Adv. Phys. X {\bf 1}, 412 (2016).}

\bibitem{X.Fan2018}
X. Fan, D. Ma, B. Fu, C.-C. Liu, and Y. Yao,\Doi{10.1103/PhysRevB.98.195437}{Phys. Rev. Lett. {\bf 98}, 195437 (2018).}


\bibitem{bzare2016}
B. Zare Rameshti, R. Asgari,\Doi{10.1103/PhysRevB.94.205401}{Phys. Rev. B {\bf 94}, 205401 (2016)}.

\bibitem{Vyborny}
K.V\'{y}born\'{y}, A. A. Kovalev, J. Sinova, and T. Jungwirth, \Doi{10.1103/PhysRevB.79.045427}{Phys. Rev. B {\bf 79}, 045427 (2009)}.

\bibitem{Faridi} A. Faridi, R. Asgari and A. Langari, \Doi{10.1103/PhysRevB.93.235306}{Phys. Rev. B {\bf 93}, 235306 (2016)}.

\bibitem{moslem-bp2}
M. Zare, B. Z. Rameshti, F. G. ghamsari, R. Asgari., \Doi{10.1103/PhysRevB.95.045422}{Phys. Rev. B {\bf 95}, 045422 (2017).}

\bibitem{Rushforth:2007_a}
A.W. Rushforth, K. Výborný, C.S. King, K.W. Edmonds, R.P. Campion, C.T. Foxon, J. Wunderlich, A.C. Irvine, V. Novák, K. Olejník, A.A. Kovalev, Jairo Sinova, T. Jungwirth, B.L. Gallagher.,  \Doi{10.1016/j.jmmm.2008.04.070}{J. Magn. Magn. Mat. {\bf 321}, 1001 (2009).}

\bibitem{Rushforth:2007_b}
A. W. Rushforth, K. Výborný, C. S. King, K. W. Edmonds, R. P. Campion, C. T. Foxon, J. Wunderlich, A. C. Irvine, P. Vašek, V. Novák, K. Olejník, Jairo Sinova, T. Jungwirth, and B. L. Gallagher, \Doi{10.1103/PhysRevLett.99.147207}{Phys. Rev. Lett. {\bf 99}, 147207 (2007).}

\bibitem{ando}
T. Ando, A. B. Fowler, F. Stern, \Doi{10.1103/RevModPhys.54.437}{Rev. Mod. Phys. {\bf 54}, 437 (1982)}.

\bibitem{sarma}
S. Das Sarma, S. Adam, E. H. Hwang, and E. Rossi, \Doi{10.1103/RevModPhys.83.407}{Rev. Mod. Phys. {\bf 83}, 407 (2011)}.

\bibitem{Liuprb16}
Y. Liu, T. Low, and P. P. Ruden,\Doi{10.1103/PhysRevB.93.165402}{ Phys. Rev. B {\bf 93}, 165402 (2016).}

\bibitem{Luonatcom15}
Z. Luo, J. Maassen, Y. Deng, Y. Du, R. P. Garrelts, M. S. Lundstrom, P. D. Ye, X. Xu.,\Doi{10.1038/ncomms9572}{Nat. Comm. {\bf 6} 8572 (2015).}

\bibitem{Xianatcom14}
F. Xia, H. Wang, Y. Jia ,\Doi{10.1038/ncomms5458}{Nat. Comm., {\bf 5} 4458 (2014).}

\bibitem{T.Cheng17}
T. Cheng, H. Lang, Z. Li,  Z. Liu  and Z. Liu, \Doi{10.1039/C7CP03736H}{Phys. Chem. Chem. Phys., 19, 23942 (2017).}


\bibitem{Sun18}
J. Sun, Y. Zhang, J. Leng,, H. Ma,\Doi{10.1016/j.physe.2017.11.012}{ Physica E. {\bf 97}, 170 (2018).}

\bibitem{Klos}
J. W. Kłos and I. V. Zozoulenko,\Doi{10.1103/PhysRevB.82.081414}{ Phys.  Rev.  B {\bf 82},  081414(R)  (2010).}

\bibitem{SarmaRev}
S. D. Sarma , S. Adam, E. H. Hwang, E. Rossi,\Doi{10.1103/RevModPhys.83.407}{ Rev Mod Phys, {\bf 83} 407–70 (2011).}

\bibitem{moslem-stt}
M. Zare, L. Majidi, R. Asgari, \Doi{10.1103/PhysRevB.95.115426}{Phys. Rev. B {\bf 95}, 115426 (2017).}

\bibitem{Faghaninia}
R. Fei, A. Faghaninia, R. Soklaski, J.-A. Yan, C. Lo and L. Yang., \Doi{10.1021/nl502865s}{Nano Lett. {\bf 14}, 6393-6399 (2014)}.

\bibitem{Ziman}
J. M. Ziman, Electrons and Phonons (Oxford University, London, 1960),

\bibitem{Gruneisen}
E. Gruneisen,\Doi{10.1002/andp.19334080504}{ Ann. Phys. (Leipzig) {\bf 16}, 530 (1933).}

\bibitem{Bloch30}
F. Bloch,\Doi{10.1007/BF01341426}{ Z. Phys. {\bf 59}, 208 (1930).}

\bibitem{Zhang-Angew18}
J. Zhang, J. Zhang, C. Cheng, J. Liu, J. Lischner, F. Giustino, and S. Meng,\Doi{10.1002/anie.201800087}{ Angew. Chem. Int. Ed. 57, 17, 1548 p 4585 (2018).}

\bibitem{Kukkonen49}
C. A. Kukkonen,  \Doi{10.1103/PhysRevB.18.1849}{Phys.  Rev.  B  {\bf 18}, 1849 (1978).}

\bibitem{Hwang08prl}
E. H. Hwang and S. Das Sarma,\Doi{10.1103/PhysRevB.77.115449}{ Phys. Rev. B {\bf 77}, 115449 (2008).}

\bibitem{Efetov10}
D. K. Efetov, P. Kim,\Doi{10.1103/PhysRevLett.105.256805}{  Phys. Rev. Lett. {\bf 105}, 256805 (2010).}

\bibitem{Toheiphon}
T. Tohei, A. Kuwabara, F. Oba and I. Tanaka,\Doi{10.1103/PhysRevB.73.064304}{ Physical Review B {\bf 73}, (6), 064304 (2006).}

\bibitem{H.Zhounpj}
H. Zhou, Y. Cai, G. Zhang and Y.-W. Zhang,\Doi{10.1038/s41699-017-0018-2}{ npj. 2D Mate. Appl. {\bf 1}, 14 (2017).}

\bibitem{PengRSC16}
B. Peng, H. Zhang, H. Shao, Y. Xu, X. Zhang and H. Zhu,\Doi{10.1039/C5RA19747C}{ RSC Adv. {\bf 6}, 5767 (2016).}

\bibitem{JainSci15}
A. Jain and A. J. H. McGaughey,\Doi{10.1038/srep08501}{ Sci. Rep. {\bf 5}, 8501 (2015).}

\bibitem{Omini}
M. Omini, and A. Sparavigna,\Doi{10.1103/PhysRevB.53.9064}{ Phys. Rev. B  {\bf 53}, 9064 (1996).}

\bibitem{Broido}
D. Broido, A. Ward, and N. Mingo,\Doi{10.1103/PhysRevB.72.014308}{ Phys. Rev. B {\bf 72}, 014308 (2005).}

\bibitem{Fugallo13}.
G. Fugallo, M. Lazzeri, L. Paulatto, and F. Mauri,\Doi{10.1103/PhysRevB.88.045430}{ Phys. Rev. B {\bf 88}, 045430 (2013).}

\bibitem{Fugallo14}.
G. Fugallo, A. Cepellotti, L. Paulatto, M. Lazzeri, N. Marzari, and F. Mauri,\Doi{10.1021/nl502059f}{ Nano Lett. {\bf 14}, 6109 (2014).}

\bibitem{moslem-physicac}
L. Majidi, M. Zare, R. Asgari., \Doi{10.1016/j.physc.2018.03.003}{Physica  C {\bf 549}, 77 (2018).}

\bibitem{Wangpccp16}
Z. Wang, T.-Y. Lü, H.-Q. Wang, Y. P. Feng, J.-C. Zheng., \Doi{10.1039/C6CP06164H}{Phys. Chem. Chem. Phys. {\bf 18}, 31424 (2016).}

\end{thebibliography}
\end{document}